\documentclass{emulateapj}

\usepackage{graphics}

\usepackage{epsfig}

\input epsf

\slugcomment{To appear in AJ}

\shorttitle{Mass Distribuition}
\shortauthors{Plana et al.}

\begin{document}


\title{Mass Distribution in Hickson Compact Groups of Galaxies}

\author{H. Plana \altaffilmark{1,2}, P. Amram \altaffilmark{2}, C. Mendes de Oliveira
\altaffilmark{3},  C. Balkowski \altaffilmark{4}}

\altaffiltext{1}{Laboratorio de Astrof\'\i sica Te\'orica e Observacional, Universidade Estadual de Santa Cruz, Rodovia Ilh\'eus - Itabuna km 16 45650-000 Ilh\'eus BA - Brazil}

\altaffiltext{2}{Laboratoire d'Astrophysique de Marseille,
Universit\'e de Provence, CNRS,
38, rue Fr\'ed\'eric Joliot-Curie
13388 Marseille cedex 13 France}

\altaffiltext{3}{Universidade de S\~ao Paulo Instituto de
Astronomia, Geof\'{\i}sica e Ci\^encias Atmosf\'ericas,
Departamento de Astronomia, Rua do Mat\~ao 1226 - Cidade
Universit\'aria 05508-900 S\~ao Paulo SP - Brazil}

\altaffiltext{4}{Observatoire de Paris, GEPI Universit\'e Paris 7, CNRS, 5 Place Jules Janssen, F-92195 Meudon
Cedex, France}


\begin{abstract}

This study presents the mass distribution for a sample of 18 late-type galaxies in nine Hickson Compact Groups. We used rotation curves from high resolution 2D velocity fields of Fabry-Perot observations and J-band photometry from the 2MASS survey, in order to determine the dark halo and the visible matter distributions.  
The study compares two halo density profile, an isothermal core-like
distribution and a cuspy one. We also compare their visible and dark matter distributions with those of galaxies belonging to cluster and field galaxies coming from two samples: 40 cluster galaxies of \citet{Bar04} and 35 field galaxies of \citet{Spa08}.

The central halo surface density is found
to be constant with respect to the total absolute magnitude similar to what is found
for the isolated galaxies. 
This suggests that the
halo density is independent to galaxy type and environment. 
 We have found that core-like density
profiles fit better the rotation curves than cuspy-like ones. No
major differences have been found between field, cluster and
compact group galaxies with respect to their dark halo density
profiles.

\end{abstract}

\keywords{galaxies: halos --- galaxies: kinematics and dynamics --- galaxies: general}

\section{Introduction}

Hickson Compact Groups of Galaxies (HCGs) are collections of three
to seven galaxies inside a three magnitude interval, where the
members have a projected separation of the order of the galaxy
diameters and a low velocity dispersion ($\sim$200 km.s$^{-1}$, Hickson
1992).  These galaxies, located in the denser environments of the
local universe, experience a high rate of galaxy-galaxy
interactions and merging. They constitute a privileged laboratory
to study different stages of interactions from violent merging to
systems with no apparent signs of interaction.

 \citet{Rub91} analyzed rotation curves for
32 Hickson compact group galaxies and found that two-thirds of
them had peculiar rotation curves. For the sub-sample of galaxies
for which rotation curves could be derived, they found
that spiral galaxies in compact groups have low mass-to-light
ratios compared to field galaxies by about 30\%, which
could be explained if compact group galaxies have smaller dark
halos than their field counterparts. Given that compact groups are
environments where tidal encounters are common, it may be expected
that interactions have stripped or disrupted the galaxy dark halos
at some level. These conclusions have important consequences for
the determination of group lifetimes and understanding how compact
groups evolve and eventually merge.

In order to address these questions, we have launched an
observational program to obtain and analyze 2D velocity fields of
HCGs using scanning Fabry-Perot techniques. Fabry-Perot imaging of
about 30 groups (about 100 galaxies) have be obtained to date.
Specific groups have been studied such as HCG\ 16 \citep{Men98}, HCG\ 92 \citep{Pla99,Men01}, HCG\ 31\citep{Amr04,Amr07}, HCG\ 90 \citep{Pla98}, HCG\ 18 \citep{Pla00} and HCG\ 79 \citep{Dur08}.
HCG\ 07, HCG\ 10, HCG\ 19, HCG\ 87, HCG\ 91 and HCG\ 96 have been studied by \citet{Amr03} and \citet{Pla03} studied HCG\ 88, HCG\ 89, HCG\ 100. \citet{Tor09} studied the kinematics and the star formation  in the intragroup medium in HCG\ 02, HCG\ 22 and HCG\ 23. 
A determination of a group evolutionary sequence was done by 
\citet{Pla02} and the study of the Tully-Fisher (T-F) relation of
the galaxies in these dense environments was done by \citet{Men03}.
They have shown, for a sample of 25 rotation curves of
late-type galaxies, that HCG galaxies follow the (T-F) relation but a
few galaxies are brighter than expected from the relation with
their mass or, alternatively, present a mass that is too low for
their luminosities.  \citet{Men03} favor a scenario in which these
outliers have been brightened because of either enhanced star
formation or merging. Alternatively, the CG galaxies may have
undergone truncation of their dark halo due to interactions. The
fact that the B-band T-F relation is similar for compact group and
field galaxies tells us that these galaxies show common
mass-to-size relations and that the halos of compact group
galaxies have not been significantly stripped inside R$_{25}$.

In this paper, we address the question of the mass distribution
(luminous and dark) of spiral galaxies in compact groups in order
to determine whether galaxies in dense environments have
experienced a different history than galaxies in less dense
environments. This work extends the study by \citet{Men03} on the
dynamics of galaxies in compact groups by investigating the
distribution of the mass, taking into account the contribution of
the bulge, the disk and the dark halo. To quantity the mass
distribution of galaxies we have used mass models based on H$\alpha$
rotation curves and surface brightness profiles. We could not derive extended rotation curves, with combined H$\alpha$ and 
HI data, as done by  \citet{Spa08} (hereafter S08) for galaxies in looser environments, given that 
there are no HI rotation curves available for the compact group galaxies in our sample.
Different profiles have been tested to model dark halos. The mass-to-light
fraction of the disk is computed together with the dark halo
parameters using the best fit model (hereafter BFM) and the maximum disk model (hereafter MDM). 
The MDM approach uses the exponential disk in order to fit the rotation curve. With the
MDM approch we use the ISO hypothesis to model the halo.

The paper is structured as follows. In section 2, we present the
data and models (photometric and kinematic). In section 3, we
review the mass distribution of galaxies in the field and in
clusters. In section 4 we present, analyze and discuss our results
on HCGs and compare them to results in less dense environments. Finally in section 5, some concluding remarks are given.
In appendix (A) we describe individually  the RCs for each galaxy and in appendix (B), from figure \ref{Figure04} to figure \ref{Figure21}, we present the J-band surface brightness profile and the RCs for
the galaxies studies.

\section{Data and models}


We have used the 2MASS survey  \citep{Skr06} to compute the
photometric surface brightness profiles. Images from the 2MASS
survey are already calibrated with a given zero point for each
field and with a pixel size of 1\arcsec~  on the sky. This pixel
size is well matched to the pixel size of the Fabry-Perot maps of
0.91\arcsec~ \citep{Amr03, Pla03}.

We have built surface brightness profiles using the J-band because
it probes the dominant stellar component out to the outskirts. 

We have selected a sub-sample of 18 galaxies among the 25 galaxies
used in the \citet{Men03} study. Indeed, 2MASS data are not exploitable for
seven of them: four galaxies (HCG19c, HCG79d, HCG89d, HCG96d) show
J-images that are too weak to be exploited and three others (HCG100c,
HCG100d, HCG96a) prove to be too difficult to derive satisfactory
surface brightness profiles.

Surface brightness profiles have been built by fitting ellipses to
the isophotes of the J-band images using the {\it ELLIPSE task} of
the {\it SDSDAS package} with IRAF \footnote{IRAF is distributed
by the National Optical Astronomy Observatory, which is operated
by the Association of Universities for Research in Astronomy
(AURA) under cooperative agreement with the National Science
Foundation}. We only fixed the center of the ellipse in the
fitting parameters and we usually used a 0.2\arcsec~ space between
two ellipses. 
 In a few cases (for HCG07c, HCG16c,
HCG19a, HCG88d), we had to extrapolate the outer profile using an
exponential disk profile in order to reach the same extension
as the rotation curves since the sky background contaminates the surface
brightness profile.

In order to use the mass model, we have performed a decomposition
of the surface brightness profile into classical two components:
an exponential disk
\begin{equation}
\mu = \mu _0 + 1.0857 \frac{r}{h}
\end{equation}

 where $\mu _0$ is the central surface brightness and $h$ is the disk scale parameter, 

 and a bulge with an r$^{1/4}$ law
\begin{equation}
\mu = \mu _e + 8.3268 \left [ (\frac{r}{r_e})^{1/4} + (\frac{r}{r_t})^4 \right ]
\end{equation}
where $\mu _e$ is the central surface brightness, and $r_e$ and $r_t$ are scale factors.

To perform the decomposition we used a home-made program based on
a minimized $\chi ^2$  routine from the MINUIT package \citep{Fle70}. The program first fits the disk using a
visual estimate of the profile. It then subtracts the fitted disk
and adjusts the bulge. The operation is repeated in order to
minimize the $\chi ^2$ for each disk and bulge parameter. Using
this method, the structures seen in the disk surface brightness
profiles are carried on to the mass profiles. We then have five
parameters, two for the disk ($\mu _0$ and $h$) and three for the
bulge ($\mu _e$, $r_e$ and $r_t$). Table 1 shows the results of the fit.
We also list the values of inclination used to derive the rotation curves, from velocity fields, published in 
\citet{Amr03}, \citet{Pla03} and \citet{Men98}.

In order to compute the mass distribution, we have combined the
stellar light contribution derived from the photometry with the
dark halo contribution from the rotation curves using the
model developed by \citet{Car85} and revised by \citet{Bla01}.
The surface brightness profile is transformed into a mass
distribution for the stellar disk and the stellar bulge assuming a
variable but radially constant mass-to-light ratio M/L
\citep{Cas83, Car85}. Mass decompositions into luminous and dark
components are not unique, mainly due to the uncertainties of the
stellar disk M/L \citep{Bar04} but also due to our ignorance
 of the actual dark halo density profile. We
minimized this problem by using J-band photometry in order to
narrow down the stellar component to the old stellar populations.
This mass model estimates the contribution of each velocity
component (bulge, disk and halo) using the photometric profile
decomposition for the luminous contribution (bulge and disk) and a
predetermined density profile for the halo.

The luminosity profile, in the J-band, is used to probe the
mass-dominant stellar component and this is transformed into a mass
distribution for the stellar disk assuming a variable but radially
constant M/L ratio for the disk and the bulge.  
For this study we used two different density
distributions to represent the dark halo. First we used a
spherical distribution given by an isothermal sphere with a
density profile given by:

\begin{equation}
\rho = \frac{\rho _0}{[1 + (\frac{r}{r_0})^2]^{3/2}}
\end{equation}

where $\rho_0$ is the central halo density and $r_0$ the core radius. 
We also used the Navarro, Frenk and White \citep{Nav96} (hereafter NFW) density profile:

\begin{equation}
\rho = \frac{\rho _0}{\frac{r}{r_0}(1 + \frac{r}{r_0})^2}
\end{equation}

The result of the quadratic sum of each velocity component (bulge, disk
and halo) is fitted to the observational rotation curve by minimizing the
reduced \footnote{The reduced $\chi^2$ is the $\chi^2$ divided
by the number of degrees of freedom.} $\chi ^2$ in the four dimensions
space: $(M/L)_{disk}$, $(M/L)_{bulge}$, $\rho _0$ and $r_0$.
The reduced $\chi^2$ can be affected by small and large scale variations in
the measured velocities that can have various origins (bars, other
non-aximmetric motions, star formation, projection effects, etc)
that the models does not take into account. A high $\chi^2$ does not therefore
necessarily means a bad fit. A reduced $\chi^2$ around 1 is an indication
of a good fit, but because of the degeneracies between parameters the
result may not be unique. We looked if errors listed in table 1 and 2 for the halo and M/L parameters are correlated with the reduced $\chi^2$. It appears that the reduced $\chi^2$ remains more or less constant with the errors, showing no correlations.
In appendix (A) we give comments on each galaxy analysis. In appendix (B) (from figure \ref{Figure04} to figure \ref{Figure21}) we present the fits to
the J-band surface brightness profiles and the rotation curves
using the isothermal sphere to model the dark halo. The velocity uncertainties, presented on the plots, are the differences between the two sides
of the rotation curve (receeding and approaching).
In appendix (A) we also give a description of the fit for both
the NFW and maximum disk hypothesis
for the dark halo, but the corresponding plots are not presented
in the paper.

\section{Comparison samples of field and cluster galaxies}

The main goal of this work is to compare the mass
distributions for galaxies in dense environments with the ones found
in less dense or isolated environments. Two different works can be
found in the literature to which we could compare our results:
\citet{Bar04} (hereafter B04) studies on galaxies in clusters and \citet{Spa08} (hereafter S08) studied field galaxies. 

\subsection{Cluster Galaxies - B04}

B04 shows rotation curves for 40 spiral galaxies from the \citet{Sch93} survey, observed with a FP
interferometer. B04 followed a
similar technique to ours using both the photometry (in the
I-band) and the RCs in order to fit the halo parameters and
determine the mass distribution. The photometric profiles used by
B04 come from \citet{Pal00}. Two dark halo models were used by B04, same as those used by us, a pseudo-isothermal and the NFW
density profile. The sample from \citet{Sch93} is not very
homogenous with respect to galaxy environment. More than half the galaxies
present comes from clusters. In order to compare our results with
cluster galaxies, we had  to convert the NFW dark halo parameters they used
into our set of parameters. We deduced the central density
($\rho_0$) and the core radius ($r_0$) from the $v_{200}$ and $c$
from B04 using definitions found in NFW. The
parameters for the iso-thermal halo are the same as ours.
B04 found that stars-only model fits most of the RCs of their sample over a significant radial range. But the addition of
a dark halo generally leads to low values of M/Ls. They also confirm a scaling law between the central halo density and the core radius.
\citet{Pal00} used I-band photometry and the $M/L$
indicated in B04 were determined in the same band. In
order to compare the results from our sample with the cluster galaxies, it was
necessary to transform our surface brightness profile from the
J-band to the I-band. We first made the hypothesis that the
profile in the J-band has the same shape and slope as in the I-band (same
scale length for the exponential disk  and same r$_e$ for the
bulge). 

Only the $\mu_0$ and $\mu_e$ were modified. We used
the apparent magnitude found in NED \footnote{NASA/IPAC Extragalactic Database (NED) which is operated by the Jet Propulsion Laboratory, California Institute of Technology, under contract with the National Aeronautics and Space Administration}  in the SDSS i-band for our galaxies
and determined the absolute magnitude in the Johnson I-band using
\citet{Fuk95}  formula and the distances noted in the RC3 \citep{deV91}.
We scaled the surface brightness profile of our galaxies in order
to get the calculated I-band absolute magnitude using our mass model
program.  Hereafter we perform the comparison with all the cluster galaxies sample from B04 and our HCG galaxies and we discuss the results in the conclusion but we did 
not plot it  in the paper.

\subsection{Field Galaxies - S08}

We also wanted to compare with the field galaxies from the GHASP survey
(Galaxies HAlpha Survey Project, \citet{Gar02}, \citet{Gar03},
\citet{Gar04}, \citet{Gar05}, \citet{Epi08a}, \citet{Epi08b}) where late-type field galaxies have
been observed using a Fabry-Perot interferometer. S08 present the mass model for a sub-sample of 36
galaxies of the GHASP survey. For this survey, we had to convert our J-band surface brightness
profiles into R-band. For HCG 16a, HCG 88a, HCG 88c, HCG 89a and
HCG 89c, we directly  used the surface brightness profiles from
\cite{Rub91} and for HCG 96c we used the profile from
\citet{Ver97}. We estimated a mean shift between $\mu_0$ in the
J-band and $\mu_0$ in the R-band for the remaining galaxies and we
verified that the $M_R$ magnitudes obtained with the mass model
program were consistent with what we found in the litterature
\citep{Hic89}. S08 confirm the halo scaling laws showing that low luminosity galaxies have small core radius and high central halo density and that the halo surface density is constant with the absolute
magnitude. They also found that a trend can be seen between the core radius and the color or the disk scale length.

The galaxies from the cluster galaxies sample have absolute B magnitudes between: $-21.5 <
M_B < -19.0$ and the field galaxies B magnitudes are between
$-22.0 < M_B < -15.5$. Our sample of HCG have $-21.5 < M_B <
-18.5$. We performed a KS test between our sample and the cluster galaxies
sample plus a sub-sample of the field galaxies with an $M_B < -18.5$ representing 16 galaxies.
The results show they are similar at the 85\% level or higher. We
then used this sub-sample of the field galaxies sample with
 $M_B < -18.5$ to perform the comparison.

\section{Mass distribution in galaxies in compact groups: comparison with field and clusters galaxies}

In this section we discuss the relationships between halo parameters, disk M/L ratios, halo and disk masses using
the different halo models and also comparing with other galaxies samples  (figures \ref{Figure01} to \ref{Figure03}).

In order to point out differences in the halo shape parameters between galaxies in different environment, we compared the two samples, cluster and field galaxies, using both the ISO and the NFW dark halo profiles.

Thus we analyse the disk M/L and the halo mass fraction as a function of the B absolute magnitude (see figure \ref{Figure2}) and we compare the halo mass fraction and the disk M/L using a ISO or a NFW halo (see figure \ref{Figure03}). 


We only compare with field galaxies and we derived the M/L within the R-band. 
For the halo central density and the core radius, we estimated the errors of the fit by taking the 1$\sigma$ error of the $\chi ^2$ minimization. We plot the maximum error bars determined for  different figures in order not to overfill the plots.

Finally we show differences between the central halo density and core radius using ISO and NFW halo models and we compare the $\chi^2$ for those two halo models (figure \ref{Figure03}).

\subsection{Halo parameters}

In order to make a fair comparison of the halo parameters, we evaluated these parameters using the R and I photometry for the HCG galaxies. 
Coefficients of the linear regression are quite similar and differences are inside the error bars of the linear fit. For this reason
we present in figure \ref{Figure01}, the halo parameters derived using the J-band photometry.

\subsubsection{Central halo density versus core radius ($\rho_0~ vs~ r_0$)}

Figures \ref{Figure01}a,b show the tight correlation between the central halo density $\rho_0$ and the core radius $r_0$ using the ISO and NFW models for galaxies in three different environments.
Figure \ref{Figure01}a shows the strong  correlation between the central density and the core radius for all samples using the ISO model. 
Table 3 summarises coefficients of the linear regression with the correspondent correlation coefficients.









These fits were obtained excluding 5 galaxies from the cluster galaxies sample because they were several orders of magnitude out of range. For the cluster galaxies sample, 5 galaxies show highly peaked halos in comparison with the other galaxies. Cluster galaxies galaxies show that part of the sample has lower core radii, when compared to HCG and field galaxies. This is mainly due to the fact that B04 uses a pseudo-isothermal halo density profile instead a strict isothermal model.  With the MDM model (not shown on the plot), the relation is also very well defined with a correlation of:

$Corr = -0.70 \pm 0.26$ (see table 3).

Figure \ref{Figure01}b shows the same correlation as before but using NFW model. It is clear that the relation between the central halo density and the core radius is much scattered using the NFW than the ISO model.  






The correlation found using the NFW model is $ Corr = -0.48  \pm 0.46 $ for the HCG galaxies and the central density of the regression is lower than the ISO model.
The correlation between the central halo density and the core radius is very strong for the three samples using ISO or NFW models. The correlation
coefficient is almost the same for the cluster galaxies sample using either the ISO or the NFW models and quite different for the field and HCG samples using these two models.

\subsubsection{Disk scale length and halo core radius versus  absolute B magnitude ($h$ and $r_0$ vs $M_B$)}

Instead of plotting the disk scale length versus halo core radius as S08 and \citet{Kor04}, we present the disk scale length normalised to the halo core radius vs the absolute magnitude. The result is presented in figure \ref{Figure01}c. We used the all field galaxies sample in order to display the less luminous galaxies. For the field galaxies it is clear that the ratio is constant, showing that the core radius and the disk scale length are connected somehow and that it is independent from galaxy luminosity. The result for the three samples is consistent. We perform the plot using the ISO model only, but the result is almost the same if we use NFW.
From \citet{Fre70} and \citet{Par02} we know that the disk scale length is proportional to the galaxy luminosity for a large variety of galaxy types.  This relation is also found for field and HCG galaxies. We verified that the core radius increases with the galaxy luminosity.

\subsubsection{Halo surface density versus absolute magnitude ($\rho_0*r_0~ vs~ M_B$)}

In Figure \ref{Figure01}d, we present the halo central surface density ($M_{\odot}.pc^{-2}$) versus the B absolute magnitude for the three samples using the isothermal sphere.
HCG galaxies from our sample are brighter than the field galaxies, but the halo surface density is almost the same between the three samples using the ISO model.
S08 found, from the central halo density and the core radius as a function of the absolute magnitude, that faint galaxies show a more
concentrated dark halo than massive late type galaxies. S08 confirmed the results of \citet{Kor04}, with a slightly higher constant halo density. However, with the cluster galaxies sample and our HCG galaxies, it is difficult to confirm this trend because of the lack of faint galaxies in both samples.
The results using NFW model are very similar to what can be observed with ISO.

\cite{Rom07}  have investigated the evolution of galactic dark matter haloes through collisionless high resolution N-body simulations. They found that haloes evolve through a series of quiescient phases, well fitted by a NFW profile. They showed that the characteristic density $\rho_s$ and the radius $r_s$ are strongly correlated. 


\subsection{Disk mass-to-light ratio}

In this section, we discuss the fitted disk M/L
parameter using the ISO, the NFW models  (see figure \ref{Figure02}a and \ref{Figure03}a). 
  
The dependence of disk M/L on the
absolute magnitude for both samples (field and HCG galaxies) is analysed in figure \ref{Figure02}a using only the
ISO model. 
More  than half the HCG galaxies show a very low disk
mass-to-light ratio (inferior to one). If we limit the field galaxies sample to galaxies with
similar magnitudes ($M_B < -18.5$), we found that slightly less than
half field galaxies have disk M/L larger than 4.0 when the
maximum M/L for HCG galaxies is 3.9. 
\textit{HCG galaxies have a lower M/L disk ratio than isolated galaxies
in the same range of absolute blue magnitude.} This may be due to galaxies
in Hickson groups having luminosities boosted by star formation triggered
by interactions or their maximum velocities diminished by tidal stripping.

The disk mass-to-light ratio between ISO and NFW halo models for both the HCG galaxies and
the field galaxies of GHASP are analysed in figure \ref{Figure03}a.  We first can notice that  M/L values
for the HCG galaxies are spread over a large range.  This situation is clearer for field galaxies. Almost all galaxies show a greater M/L using the ISO model.
(a) With HCG 87a, the NFW model can't fit the rotation curve with a minimum of halo
and a little disk, while ISO puts the disk to almost zero and
extend the dark halo.  (b) For HCG 91c, NFW puts the bulge to zero and gives a
larger disk than ISO. The result is that the central part is
better fitted with ISO and the outer part of the rotation curve is
better using NFW and the $\chi^2$ is very similar for both. (c) The
situation is similar with HCG 007c because of a bump that neither
model can fit. The fact that NFW gives lower M/L disk ratios is
consistent with what B04 found for their sample. In their case, the
photometry was done in I-band, but it does not change significantly the trend.

We also compared the disk M/L for HCG between ISO and the maximum disk model (not shown here). The agreement is high even if MDM has, by definition, a natural tendency to give
an higher disk M/L. This means that the BFM gives a solution that is very close to MDM and thus MDM appears to be the favourite solution.

In conclusion, using the ISO model instead of the NFW model results in a greater disk M/L and that, at least for HCG galaxies, the disk mass is independent of the galaxy luminosity.

\subsection{Halo and disk masses}

This section presents relations between halo and disk masse as a function of absolute magnitude for two different models: ISO and NFW.  Table 4 presents the disk mass, halo mass and halo mass fraction using the different models ISO, NFW and MDM. In the determination of the total mass,  the HI mass was not taken into account.

Figure \ref{Figure02}b shows the halo mass fraction versus the absolute
magnitude for both HCG and field galaxies.
The plot shows that only three HCG galaxies, HCG 10d, HCG 96c and HCG 88a, have a halo mass fraction lower than 0.8 within the range of absolute B magnitude and 66\% of field galaxies have halo mass fraction larger than 0.8 for a larger range of B magnitude.
This can be explained by the fact that the disk is predominent.
The others galaxies (HCG and isolated galaxies) are totally dominated by the dark halo and the dark halo is not related to the luminosity of the galaxy.

We were interested in a possible relation between $M_{Disk} / L_B$ and the disk mass for HCG and field galaxies.
Figure \ref{Figure02}c shows a clear correlation between the disk $M/L_B$ and the disk mass for both HCG and field galaxies. As for all figures 
in that section, we have plot the  all sample of field galaxies, if we restrict the field galaxies sample to those with an absolute B magnitude inferior 
to -18.5, we noticied that HCG galaxies and field galaxies follow the same relation.

This is confirmed with a work by \citet{Sal08} that analyse disk masses for 18 spirals by both modelling their rotation curve and by fitting
spectral energy distributions. The authors found a good agreement between the two methods. Beside their discussion about the best
way to calculate the disk mass, they present the relationship between the stellar mass, the color and the stellar mass-to-light ratio. They found  that the M/L is increasing with the disk mass. 


Based on a technique of mass decomposition of spiral galaxies from disk kinematics \citep{Per90}, found that the ratio of the disk mass to halo mass fraction shows a strong dependency with respect to the luminosity.
Their study points out a higher disk mass compared to our sample or field galaxies. 

Figure \ref{Figure03}b compares the relation between the ratios
halo mass fraction / total mass estimated with ISO and NFW
models for both HCG and field galaxies. A KS test between the two halo mass fractions from field and compact group galaxies for the same range of absolute B magnitude, 
proved that both samples have 90\% chance to be drawn from the parent distribution, indicating that there is no statistical difference between
the two 
Four galaxies (HCG 007c, HCG 10d, HCG 87c and HCG 91c) show lower halo masses using the
NFW profile. 
The halo is much more important for HCG 007c using the ISO model with respect to NFW, but both fit shows a poor $\chi^2$.
Even if HCG 87c does not show a disturbed RC, NFW halo model provides a poor fit. No fit is possible to
mimic the rotation curve with a NFW halo model. For HCG 91c, both
models lead to almost the same amount of disk, but NFW uses 10 times
less halo than ISO for almost the same $\chi^2$.

In conclusion, for both field and HCG galaxies, the mass is dominated by the halo, as B04 showed, and no differences are found between galaxies in the field and galaxies in dense environment with respect to disk mass.

\subsection{Isothermal model versus NFW halo profile}

We have compared  the halo parameters (core radius and central density) for both models (NFW and ISO)  for both HCG and field galaxies.
Figures \ref{Figure03}c,d show that the central halo density with ISO is higher than NFW and, in the mean time, the halo core radius is smaller with ISO than with NFW. The isothermal model gives apparently a more concentrated dark halo shape than the NWF96 model. This seems in contradiction with the "natural" cuspy shape of NFW dark halo profile. Indeed ISO profile is shallow in the center while NFW is more peaked. (see \citet{Bla00} for the comparison between the two halo profiles). The different parameters of the halo $\rho_0$ and $r_0$ are derived together with the M/L of the disk using the best fit model on the RCs. 
Since the slope of the RC is not high enough to be fitted by the cuspy NFW profile, it favours the disk and thus underestimates the halo. This is not the case with the core - like ISO model which matches the relatively faint slope of the RC and thus minimizes the disk and favours the halo.   
 
Figure \ref{Figure03}e compares the quality of the fits using a $\chi^2$ parameter between the isothermal sphere and the NFW models for both HCG and field galaxies.  To display the result, we have limited the $\chi ^2$ range between 0 and 15. Two galaxies (HCG 19a and HCG 91a) show a high $\chi^2$ ($>30$) with both models. Two galaxies (HCG 087c and HCG 88d) show a very poor $\chi^2$  with the NFW model only. Three other galaxies (HCG 16c, HCG 19b and HCG 89c) show a higher $\chi^2$ for the NFW than for the isothermal sphere. S08 noted that NFW gives higher $\chi^2$ than the ISO model. Figure \ref{Figure03}e confirms this for HCG galaxies. 

We can conclude that, as for field galaxies, HCG ISO dark halo fits the RC better than the NFW model. 

\section{Summary and conclusions}

Based on a sample of $\approx$ 100 velocity fields of spiral galaxies belonging to 25 HCGs, our previous study has demonstrated that:  40\% (38 galaxies) of the HCG galaxies analysed in our sample do not allow  the computation of the RC. This is because these galaxies are strongly perturbed by interaction and mergers. 33\% (31 galaxies) of HCG galaxies are regular enough to calculate a RC but the RC is not symmetric to allow the derivation of a mass model. These perturbed galaxies are mildly interacting. The remaining 27\% (25 galaxies) of them are suitable to derive mass models. For a variety reasons only 18 galaxies have been studied  in the present work. The RCs have been combined with 2MASS J-band surface brightness profiles to derive their mass models. Two shapes of dark halos have been considered: a core-like density profile (isothermal sphere or ISO) and a cuspy-like one (NFW). Best fit models and maximum disk models have been computed for the two dark halo profiles. We have compared the HCG galaxies with two samples of galaxies in different environments: field and clusters galaxies. The three samples (HCG, isolated and cluster) have been analysed using the same tools.


The results are summarized below:

\begin{itemize}

\item No obvious differences can be found between the halo parameters for the HCG galaxies and galaxies in other environments. The strong correlation between $\rho_0$ and $r_0$ is present for the three samples and the slopes of the linear
regressions between them are very similar (between -1.62 and -1.00), well inside the uncertainties. In a study based on late-type and dwarf spheroidal galaxies, \cite{Kor04} confirms this correlation giving a slope of  -1.038. On the other hand, the linear regression constant found by \cite{Kor04} is different in comparison with those found with the three samples.
 This scale difference is may be due to the fact that \cite{Kor04} compiled his sample using several sources using different methods to obtain halo parameters.


\item The use of the NFW model gives less satisfactory results than the ISO model. As mentioned by S08, the $\chi^2$ coefficient is usually larger when using the NFW. NFW model gives a worse correlation between the halo central density and the core radius than ISO. This is also true if we use the field galaxies sample. In contrast, the NFW model gives a consistent result for the cluster galaxies sample. Halo profiles are closer to isothermal spheres than NFW profile as already found by other authors \citep{Bla01,deB02,Swa03,Kas06,Spa08}.
    The slope of the linear regression between $\rho_0$ and $r_0$ is higher for thecluster galaxies sample than for the two other samples when the NFW model is used.
The disagreement between ISO and NFW is smaller with cluster galaxies sample.  61\% of HCG galxies shows higher disk M/L using the ISO model compared to NFW.
No relation between the disk scale length and the halo central density is seen using either ISO or NFW models.

\item We explored the possible connection between the halo parameters, the halo mass fraction and the disk M/L. The halo mass is high for both field galaxies
and compact groups galaxies (75 to 95\% of the total mass), leaving only modest room for the disk mass.
The halo surface density is independent of the absolute B magnitude and no clear relation is seen between the disk M/L and $M_B$, meaning that the halo is independent of the galaxy luminosity.

\end{itemize}


Because of the lack of low luminosity HGC galaxies, we could not investigate properl the possible relation between the disk scale length and the halo core radius
as S08 and \citet{Don04}. We just can say that, for our sample of bright HCG galaxies (M$_B < -18.5$) no clear trend could be found.

\acknowledgments

HP would like to thank the Scientific Council of the Laboratoire d'Astrophysique de Marseille (LAM) for its financial support during the
visits in February 2007 and 2008. HP thanks the CAPES foundation for its financial support during HP fellowship at LAM during the year 2009 (process 3656/08-0).
HP would like to thank Dr. D. Russeil for its help in the surface brightness decomposition. HP wishes to thank Niraj Welikala for the English proof reading of this paper. 
The authors wish to thank the anonymous referee for its comments.
This research has made use of the NASA/IPAC Extragalactic Database (NED) which is operated by the Jet Propulsion Laboratory, California Institute of Technology, under contract with the National Aeronautics and Space Administration. This publication makes use of data products from the Two Micron All Sky Survey, which is a joint project of the University of Massachusetts and the Infrared Processing and Analysis Center/California Institute of Technology, funded by the National Aeronautics and Space Administration and the National Science Foundation.



\clearpage

\appendix

\section{Comments on individual galaxy}

{\bf HCG07c}: HCG07 is composed of four galaxies \citep{Hic92}. We
only analyse the kinematics of HCG07c, a galaxy with morphological
type SBc. The rotation curve rises to a velocity of about 170
km/s, at a radius of r $\sim$ 35\arcsec (10 kpc) and then goes
down to a plateau, at a lower velocity of $\sim$ 150 km/s, out to
a radius of 52\arcsec (14.5 kpc). At small radius, the curve is
not regular: a plateau is present between 5\arcsec (1 kpc) and
8\arcsec (2 kpc). Because of the irregularity of the RC, the fit
is not good, especially in the center.  The bulge component is
rather significant in the fit, in order to take into account the
first 5\arcsec (1 kpc) of the curve. The halo has, by far, the
largest contribution of the fitted RCs. It reaches a plateau at
$\sim$ 35\arcsec (10 kpc). The disk, on the other hand, is weak
and its contribution is no more than 10\% of the total rotation
velocity.

When using NFW model and the maximum disk hypothesis, we found a
more prominent disk and the halo shows a plateau around 20\arcsec
(6kpc).

The photometric profile is well fitted with a bulge and an
exponential disk. Note that the disk profile presents wiggles
(most probably due to the spiral arms). In addition, the fit of
the bulge within the first 4\arcsec (0.8 kpc) is not good, as can
be seen from the  well
visible deviations in figure 1. \\

{\bf HCG10d}: HCG10 is composed of four galaxies. We applied the
mass model to HCG10d, an Sc-type galaxy. The rotation curve rises
regularly to 149 km s$^{-1}$, out to 20\arcsec (6 kpc). No bulge
has been used to fit the RC. The fit is very good and both the
disk and the halo almost have an equal contribution in the RC. The
halo shows a plateau around 10\arcsec (3 kpc) and its contribution
is more regular than the disk. As for the NFW and maximum disk
models,  the contribution for the disk is larger than for the
isothermal halo. In particular, in the case of NFW, the disk is
responsible for almost 90\% of the velocity of the RC.

The photometric profile shows that an exponential disk alone is
enough
to fit it.  \\

{\bf HCG016a}: HCG16 is  a group formed of four galaxies. HCG16a
is an SBab type galaxy. The rotation curve is a typical example of
a late-type rotation curve with a sharp rise in the first 4\arcsec
(3 kpc) till the maximum rotation velocity of 247 km.s$^{-1}$ and
a plateau reaching out to 30\arcsec (8 kpc).  The general fit is
good (ignoring the wiggles in the profile due to the spiral arms).
The bulge has the largest contribution for any single galaxy in
our sample. The contribution of the halo and the disk are
comparable, except for the fact that the halo reaches a constant
value around $\sim$ 14\arcsec (3.5 kpc) at 140 km s$^{-1}$ and the
disk rises constantly.

The use of the NFW model for the dark halo shows a much larger
contribution for the halo (more than 65\% of the fitted RC) with a
plateau starting at 8\arcsec (2 kpc). The maximum disk model gives
a larger contribution of the disk (as expected) and the halo
represents only 30\% of the maximum rotational velocity.

The photometric profile shows that the disk profile suffers a drop
starting at 22\arcsec (5.5 kpc), below the theoretical exponential
disk. In that case, for the mass model we used a theoretical
exponential
disk from 2\arcsec (0.5 kpc) to 30\arcsec (8 kpc). \\

{\bf HCG016c}: HCG16c is an S0 galaxy. The rotation curve is
slowly rising to 200 km s$^{-1}$. The fit of the rotation curve
does not fit well an observed depression between 10\arcsec (2 kpc)
and 22\arcsec (5 kpc), but the rotation curve is almost entirely
dominated by the halo, except the first 5\arcsec (1.2 kpc), where
the disk and the bulge allow a good fit

The result is almost identical for the other two models we used.
The halo totally dominates the RC.

The photometric profile is well fitted  by a rather important
bulge and an exponential disk. In this case the wiggles are
probably not related to spiral arms but instead with the merger
nature
of this galaxy \citet{Men98}. \\

{\bf HCG19a}: HCG19 is a group formed of four galaxies and its
brightest member, HCG 19a, is an S0 galaxy. The rotation curve of
this galaxy shows a flat part within a radius of 2\arcsec (0.5
kpc) and a rise outward, reaching a velocity of 125 km s$^{-1}$ at
20\arcsec (5.5 kpc). The fit is rather poor and only the halo has
a contribution. No disk has been used to fit the RC.

The other two models do not fit the rotation curve any better. In
particular, the NFW model is a very poor fit to the data, since
it cannot reproduce the maximum velocity at 35\arcsec (9.5 kpc).

The photometric profile is well fitted with only a bulge contribution. \\

{\bf HCG19b}: HCG19b is classified as an Scd galaxy and its
rotation curve is quite irregular. It presents a small plateau
between 6\arcsec (1.8 kpc) and 12\arcsec (3 kpc). It then reaches
a maximum velocity of 90 km s$^{-1}$ at around 30\arcsec (8 kpc).
Due to these irregularities, the fit of the rotation curve is not
good.  The halo is the major contribution of the rotation curve
and the disk has only a contribution of 20\% of the total
rotational velocity.

The results of the fit using the two other models are better. In
these case the contribution of the halo is more important than
using the ISO model.

The photometric profile shows a drop at 12.5\arcsec (3.2 kpc). We
replaced the part of the profile between 12.5\arcsec (3.2 kpc) and
30
\arcsec (8 kpc) by an exponential disk.  \\

{\bf HCG87a}: HCG87 is a group formed of three galaxies, and its
brightest member, HCG87a is an Sbc galaxy. The rotation curve
shows a solid body rotation with rotational velocity rises up  to
400 km s$^{-1}$ at 25\arcsec (13 kpc). This is certainly due to
the fact that this object is almost edge on.

Fits of the RCs with the two others models show that the largest
contribution comes from the disk, but the fits are not as good as
with the isothermal halo. In both cases the halo does not
represent more than 45\% of the total rotational velocity.

The photometric profile is not well fitted by an exponential
between 12\arcsec (6 kpc) and 26\arcsec (14 kpc) where a bump is
present.

The first 7\arcsec (3.5 kpc) of the profile are well fitted by a bulge profile. \\

{\bf HCG87c}: HCG87c is classified as an Sd galaxy. The rotation
curve is regular and shows a plateau at 10\arcsec (5.5 kpc) with a
maximum velocity of 160 km s$^{-1}$. The halo fits almost
completely the RC. No bulge is present and the disk shows a
plateau beginning at 6\arcsec (3.5 kpc).

The result of the fit using the maximum disk is the same as for
the isothermal halo. On the other hand, the fit using the NFW
model is very poor and cannot reach the maximum velocity.

The photometric profile is fit only with an exponential disk. \\

{\bf HCG88a}: HCG88 is formed of four galaxies, and its brightest
member, HCG88a is classified as an Sb galaxy.  The rotation curve
rises regularly without reaching a plateau. The disk contribution
is prominent. The halo contribution rises linearly and the bulge
contribution fits well the beginning of the curve.  The fit of
this rotation curve is one of the best of our sample.

The two other models show a very good fit of the rotation curve
but the NFW model indicates a larger contribution of the halo
($\sim$ 38\%) in comparison with the isothermal halo. The disk
contribution is lower in, almost, the same proportion.

The photometric profile is very well fitted with a bulge and an
exponential disk. \\

{\bf HCG88b}: HCG88b is an SBb galaxy. The rotation curve rises
regularly. It presents a small plateau between 4\arcsec (1.5 kpc)
and 10\arcsec (4 kpc) and between 30\arcsec (11.5 kpc) and
40\arcsec (15.5 kpc). The rotation curve rises to 300 km s$^{-1}$
and then drops to 190 km s$^{-1}$. Due to these irregularities,
the "best fit" is clearly not appropriate. The largest
contribution comes from the disk. It shows a plateau at 210 kms
s$^{-1}$. The bulge also has a large contribution in order to fit
the inner part of the curve and the halo rises linearly to 190 km
s$^{-1}$.

The result of the fit using the NFW model shows a major
contribution of the halo and the result for the maximum disk is
almost identical to that of the isothermal halo.

The photometric profile is well fitted by a combination of a bulge
an exponential disk. However the disk shows some small
oscillations
which obviously cannot be reproduced with an exponential disk. \\

{\bf HCG88c}: HCG88c is classified as an Scd galaxy. The rotation
curve reaches a maximum velocity of 130 km s$^{-1}$. The fit
performed using only the halo and disk contributions is good. The
halo rises regularly and the disk profile shows an unusual shape.
The disk profile shows two peaks, at 7\arcsec (2.5 kpc) and
15\arcsec (5.8 kpc), and a small depression in between.

The maximum disk shows a similar result to that given by the
isothermal halo model. The fit with the NFW model shows that the
halo alone can almost account for the whole RC.  The disk has only
a marginal contribution.

We tested the fit of the rotation curve using a theoretical
exponential but this fit did not represent well the depression
of the rotation curve at small radii, we decide then to keep photometric profile. \\

{\bf HCG88d}: HCG88d is classified as an Sm galaxy. The rotation
curve rises regularly without showing any plateau, out to a radius
of 30\arcsec (11 kpc). The fit with a halo and an exponential disk
contributions is very good. The halo rises linearly and the disk
rises to a maximum, at 12\arcsec (4.5 kpc) and then slowly
decreases.  The maximum disk model presents similar results to
that given by the isothermal sphere, but the NFW model is not
able to fit the curve at all.

The photometric profile is very well fitted with only a disk. \\

{\bf HCG89a}: HCG89a is classified as an Sc galaxy. The rotation
curve shows a bump between 4\arcsec (2 kpc) and 8\arcsec (4 kpc)
and then the velocities rise regularly up to 220 km s$^{-1}$.

This bump can not be fitted with a larger contribution from the
bulge without spoiling the fit of the curve beyond 10\arcsec (5
kpc).

The halo profile is increasing almost linearly. The disk
contribution rises to 90 km s$^{-1} $ and then slowly decreases.
We could not fit properly the entire curve even using an
exponential disk instead of the original disk photometric profile.
The maximum disk model gives similar results even if the bump
present between 4 and 8\arcsec is not as well fitted.  The NFW
model shows that the halo alone is responsible for almost the
whole RC, but the model fails to take the bulge into account.

The photometric profile is well fitted with a bulge and an
exponential
disk, for radii up to 12\arcsec (6 kpc), after which the profile drops. \\

{\bf HCG89b}: HCG89b is an SBc galaxy. The rotation curve shows a
plateau starting at 6\arcsec (3.2 kpc), with a maximum velocity of
120 km s$^{-1}$. The halo presents the largest contribution,
showing a plateau also around 6\arcsec (3.2 kpc).  The NWF96 model
and the maximum disk basically yield the same result.

We replaced the photometric profile between 0 and 6\arcsec (3.2
kpc) and with a theoretical exponential disk, to obtain the disk
contribution of the RC. We tested the fit of the rotation curve
using the original brightness profile, but the fit is not
significantly different.

The surface brightness profile is well fitted using an exponential
disk profile except for the first 6\arcsec (3.2 kpc) where the
profile is flat. This is certainly due to an apparent double
nucleus, present
on the 2mass and DSS images. \\

{\bf HCG89c}: HCG89c is an Scd galaxy.  The fit is very good with
a larger contribution of the dark halo in comparison with the
disk.  The use of the maximum disk does not show much of a
difference in the result of the fit. The NFW fit, on the other
hand, yields the worse fit. In that case the halo occupies almost
the entire fitted RC.

As for HCG89b, the photometric profile is fitted with an
exponential
disk only, except at the very beginning, where the profile is flat. \\

{\bf HCG91a}: HCG 91 contains  four galaxies, the brightest being,
HCG 91a, classified   as an SBc galaxy. The rotation curve for
this galaxy is very irregular and, in particular, for a radius
greater than 20 \arcsec (9 kpc), the rotation curve is very
asymmetric. The major contribution comes from the bulge, the disk
contribution being much less important. The halo presents a
plateau at 10\arcsec (4 kpc). The NFW model gives an even larger
contribution from the bulge and the disk is marginal. The maximum
disk model yields similar results to those given with the
isothermal halo.

The photometric profile is well fitted with a bulge and an
exponential disk.  \\

{\bf HCG91c}: HCG91c is classified as an Sc galaxy. The rotation
curve is regular, with a plateau at 16\arcsec (7 kpc), and with a
velocity of about 120 km s$^{-1}$. The fitted rotation curve is
completely dominated by the disk. The halo only represents 30\% of
the total rotation velocity and the bulge contribution is
marginal. The fit with the NFW hardly uses a dark halo and for
the maximum disk model the result is similar to that for the
isothermal halo, only with an even higher disk contribution.

The photometric profile is well fitted by a large bulge and an
exponential disk. Since at 13.5\arcsec  (6 kpc) the profile drops
drastically, we replaced this part with a theoretical exponential
disk to derive the
disk contribution for the RC. \\

{\bf HCG96c}: HCG96c, one of the four galaxies of the compact
group HCG 96, is classified as an Sa. Its rotation curve is not
very extended. It shows a plateau at 4\arcsec (2.2 kpc), with  a
velocity of 80 km s$^{-1}$. No bulge was used, the halo rises
linearly and the disk has the largest contribution to the RC.  The
result of the fit using the maximum disk is similar and the fit
using the NFW model shows a larger contribution of the dark halo
(of about $\sim$ 15\%) in the first 4\arcsec (2 kpc). After which
the halo increases at a slower rate.

The photometric profile is well fitted with a disk and a very
small bulge, except at the outskirts of the galaxy.

\section{Surface brightness profiles and rotation curves}



\begin{figure}
\figurenum{4}

\plotone{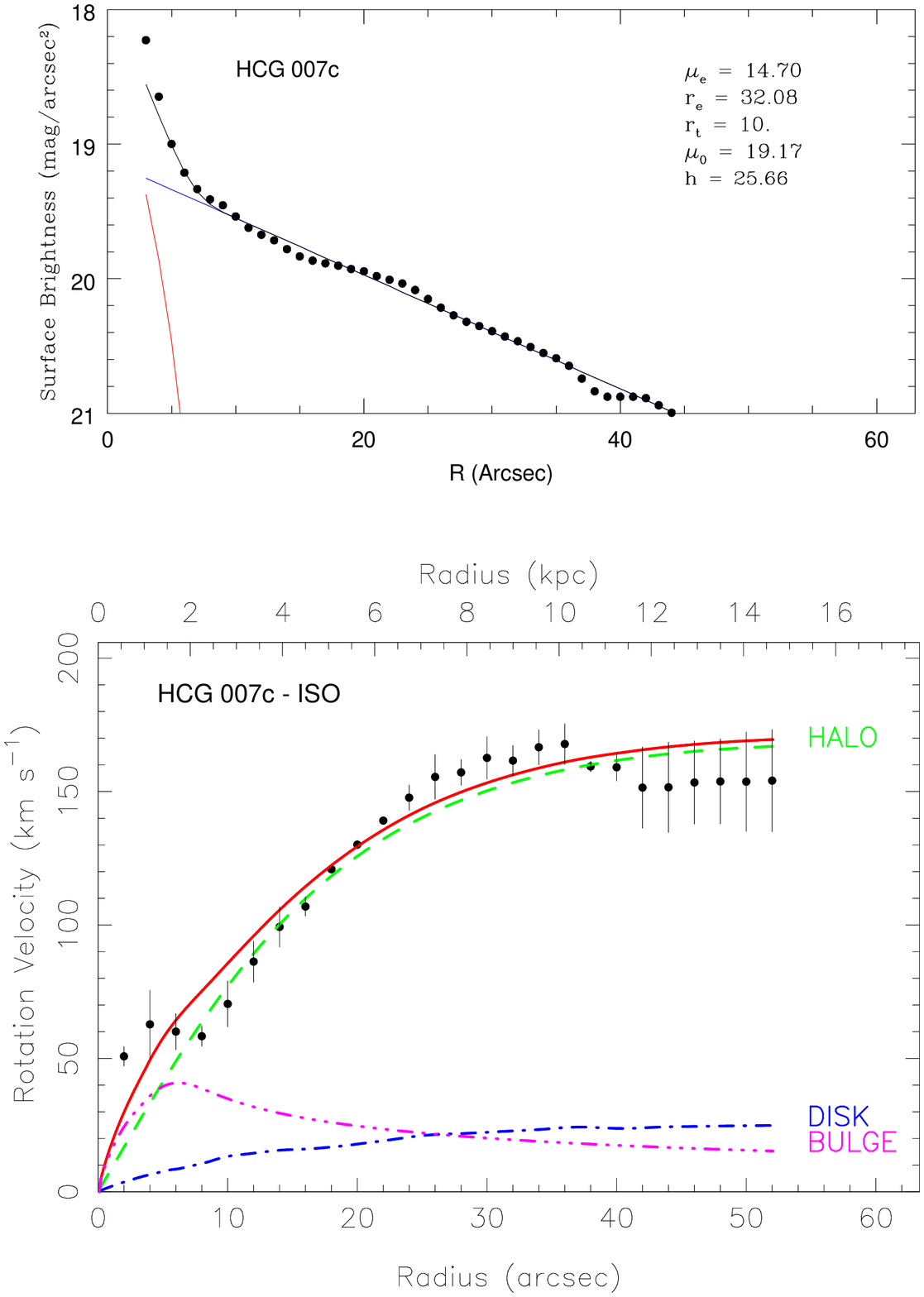}
\caption{From figure 4 to 21, we present for each galaxy the J-band surface brightness (top) and the rotation curve (bottom). Disk and bulge components are represented together with the surface brightness and the fit result. Disk component is in blue and bulge is in red on the surface brightness profile. Disk, bulge and halo velocities components are represented together with the rotation curve \label{Figure04}}

\end{figure}


\begin{figure}
\figurenum{5}
\plotone{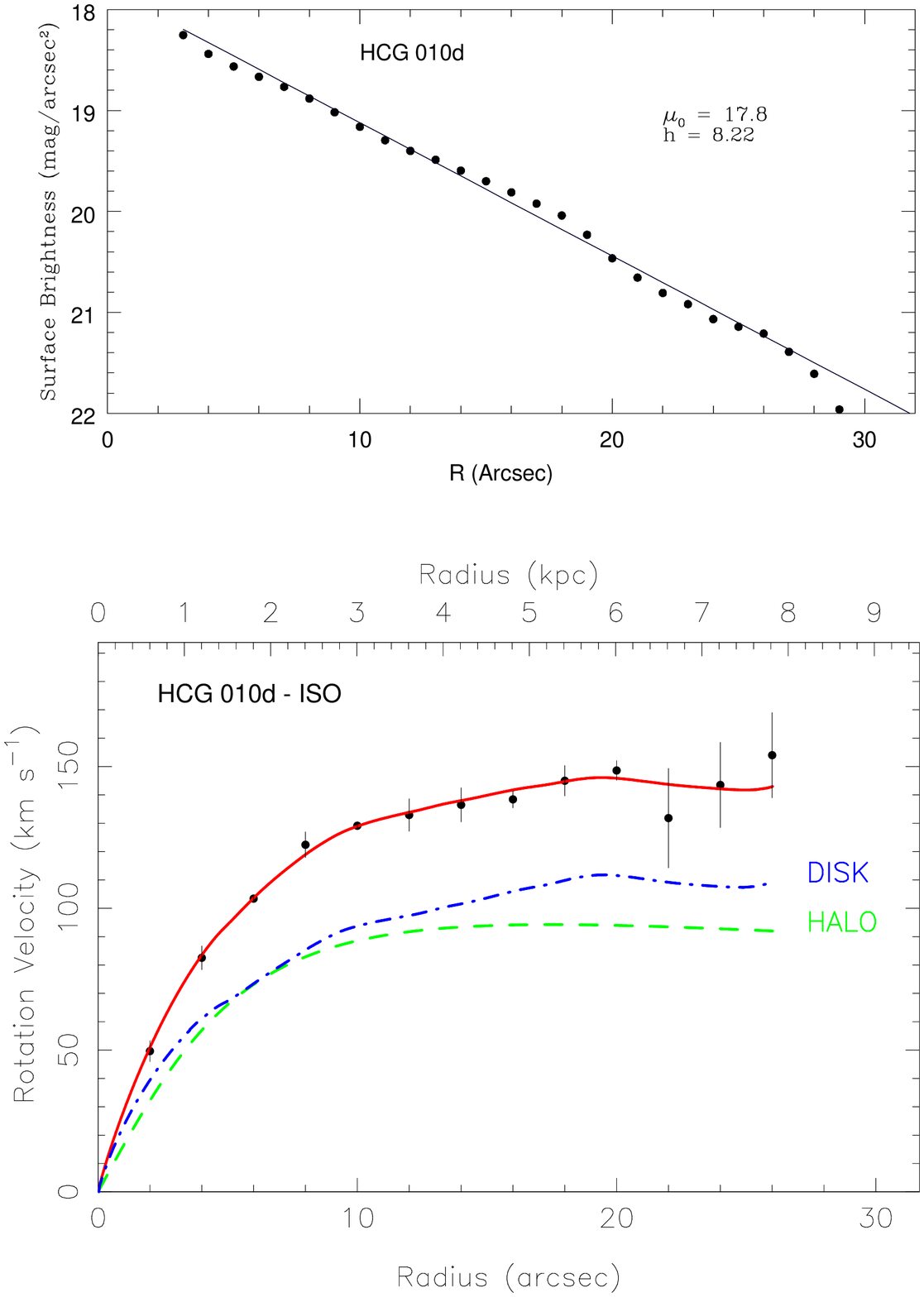}
\caption{\label{Figure05}}

\end{figure}

\clearpage

\begin{figure}
\figurenum{6}
\plotone{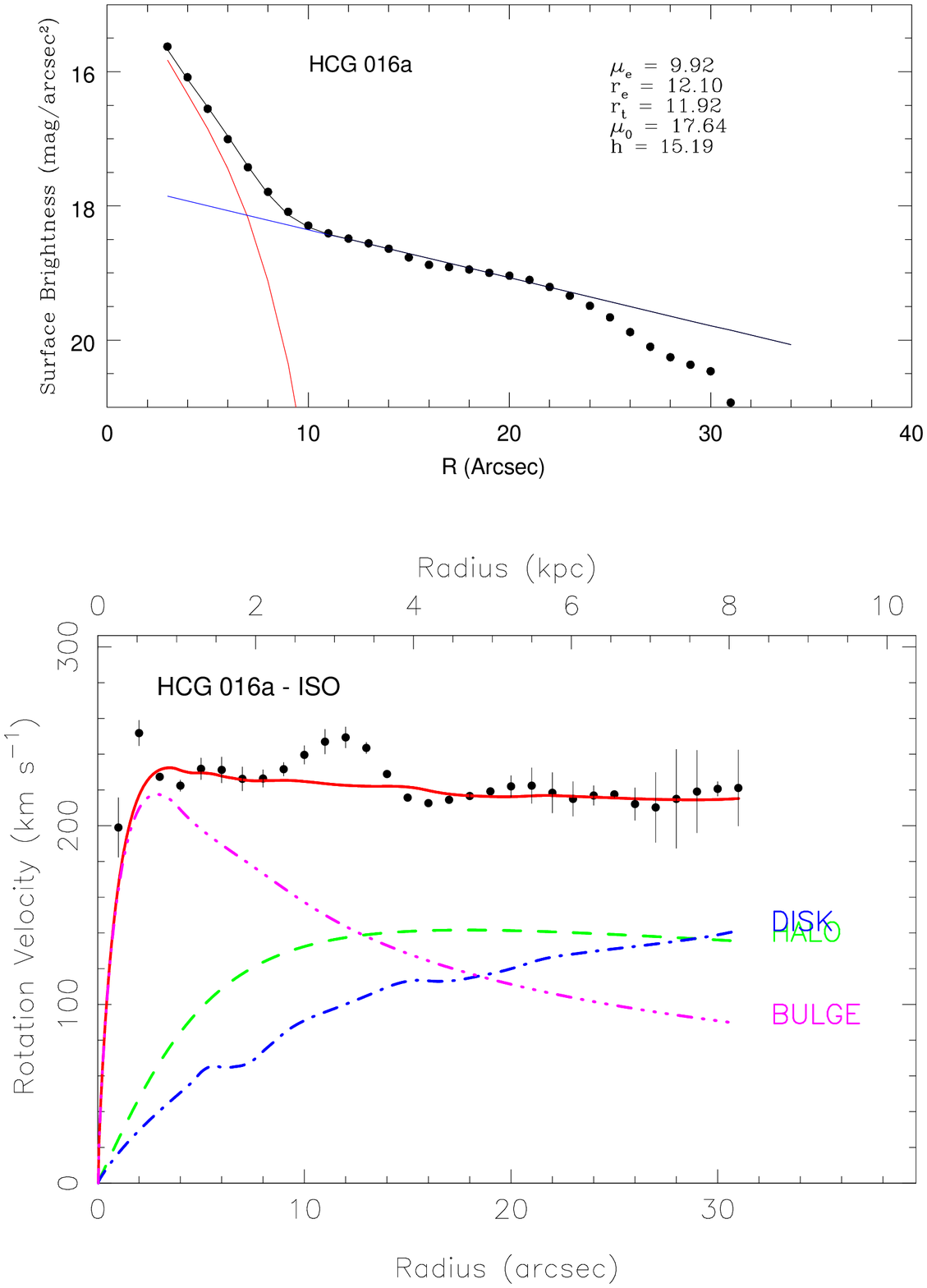}
\caption{\label{Figure06}}
\end{figure}

\clearpage

\begin{figure*}
\figurenum{7}
\plotone{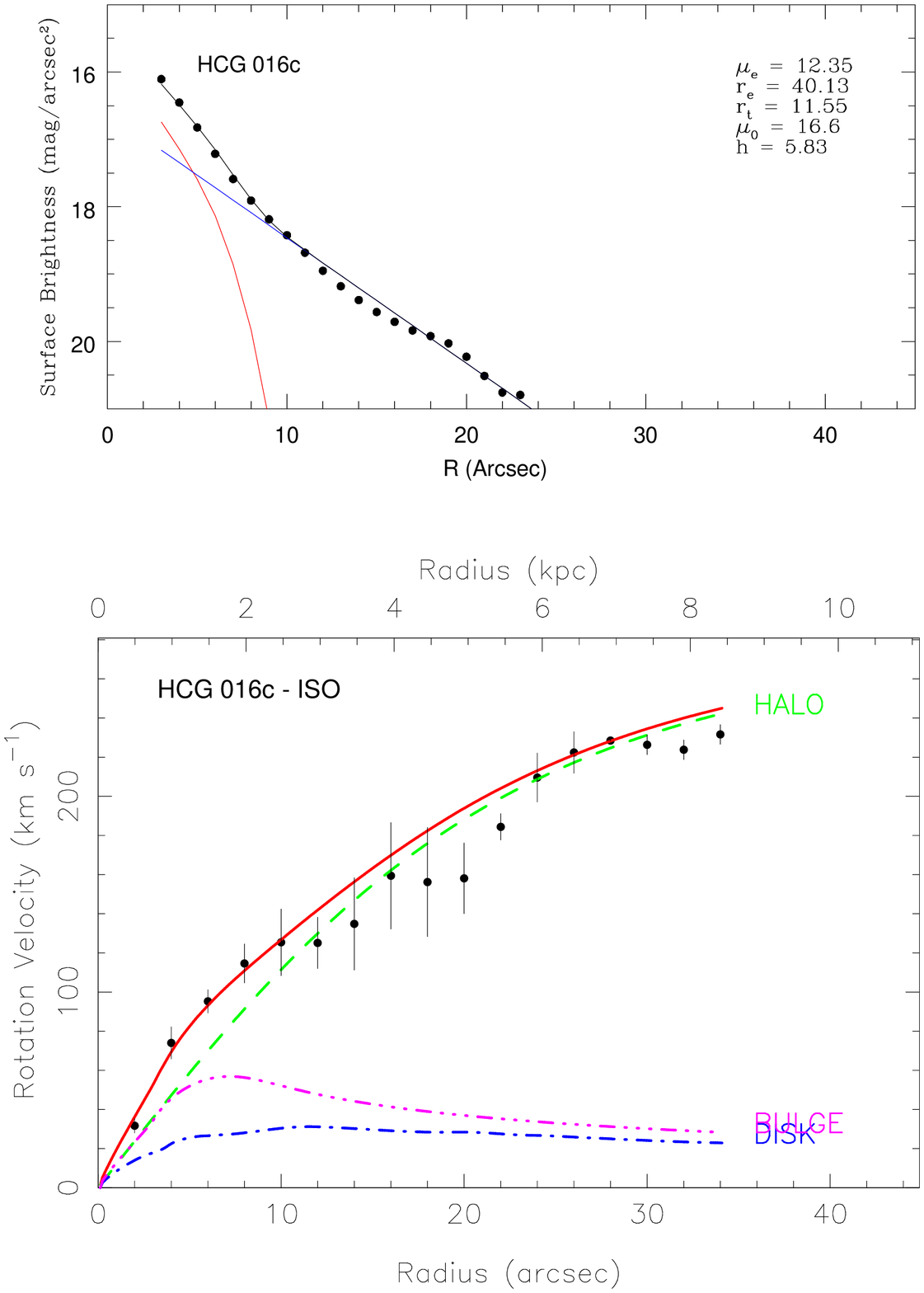}
\caption{\label{Figure07}}
\end{figure*}

\clearpage

\begin{figure*}
\figurenum{8}
\plotone{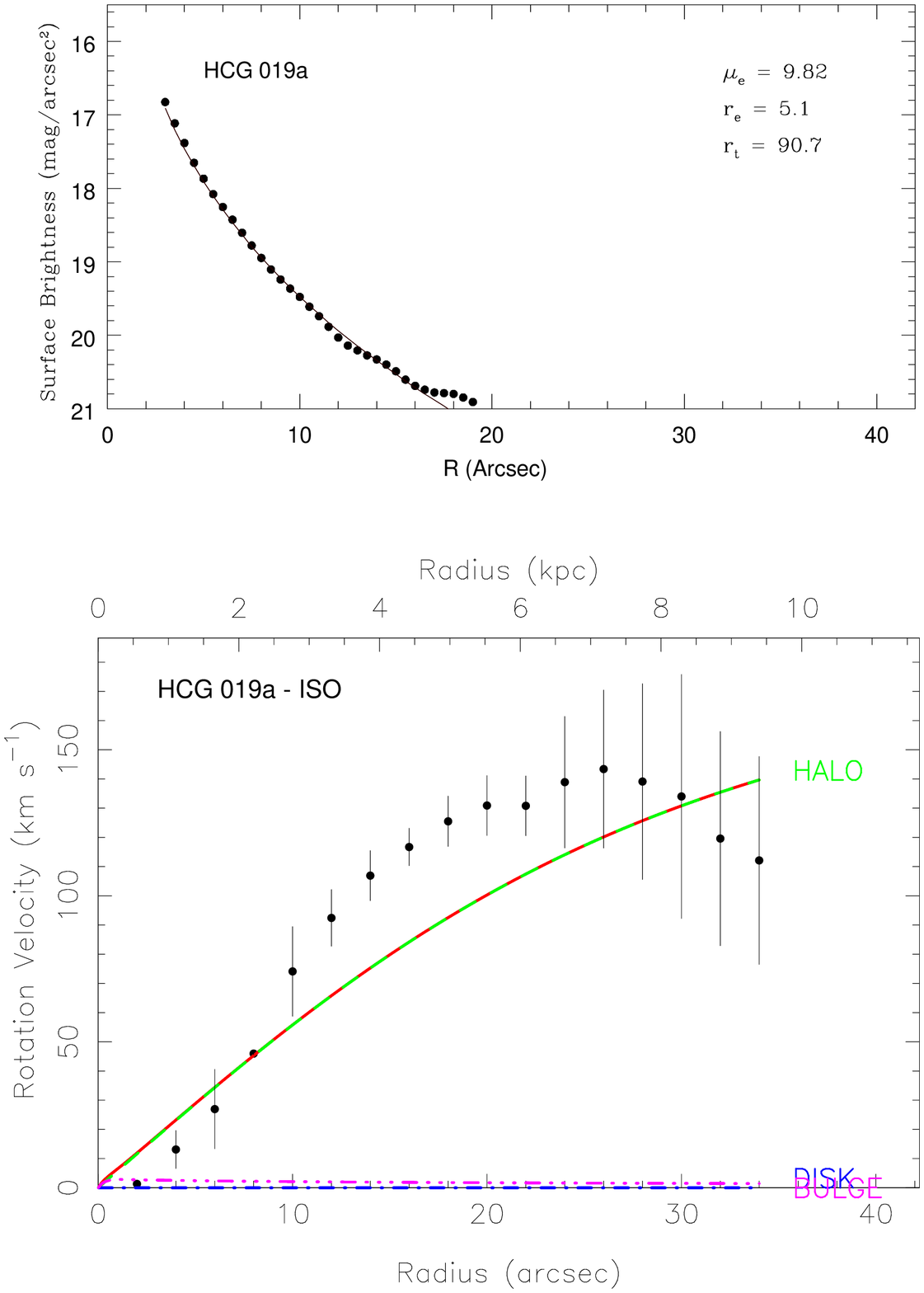}
\caption{\label{Figure08}}
\end{figure*}

\clearpage

\begin{figure*}
\figurenum{9}
\plotone{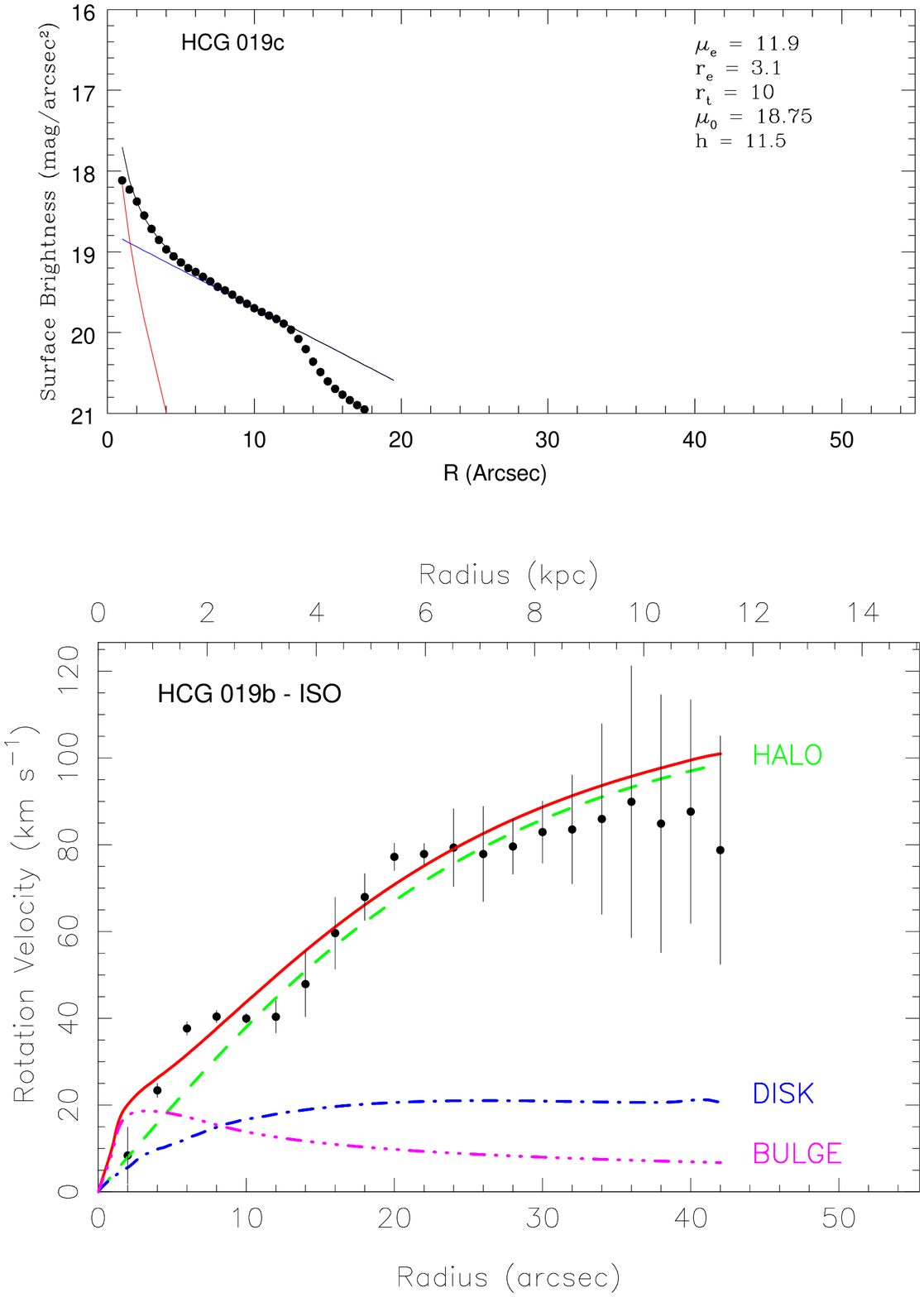}
\caption{\label{Figure09}}
\end{figure*}

\clearpage

\begin{figure*}
\figurenum{10}
\plotone{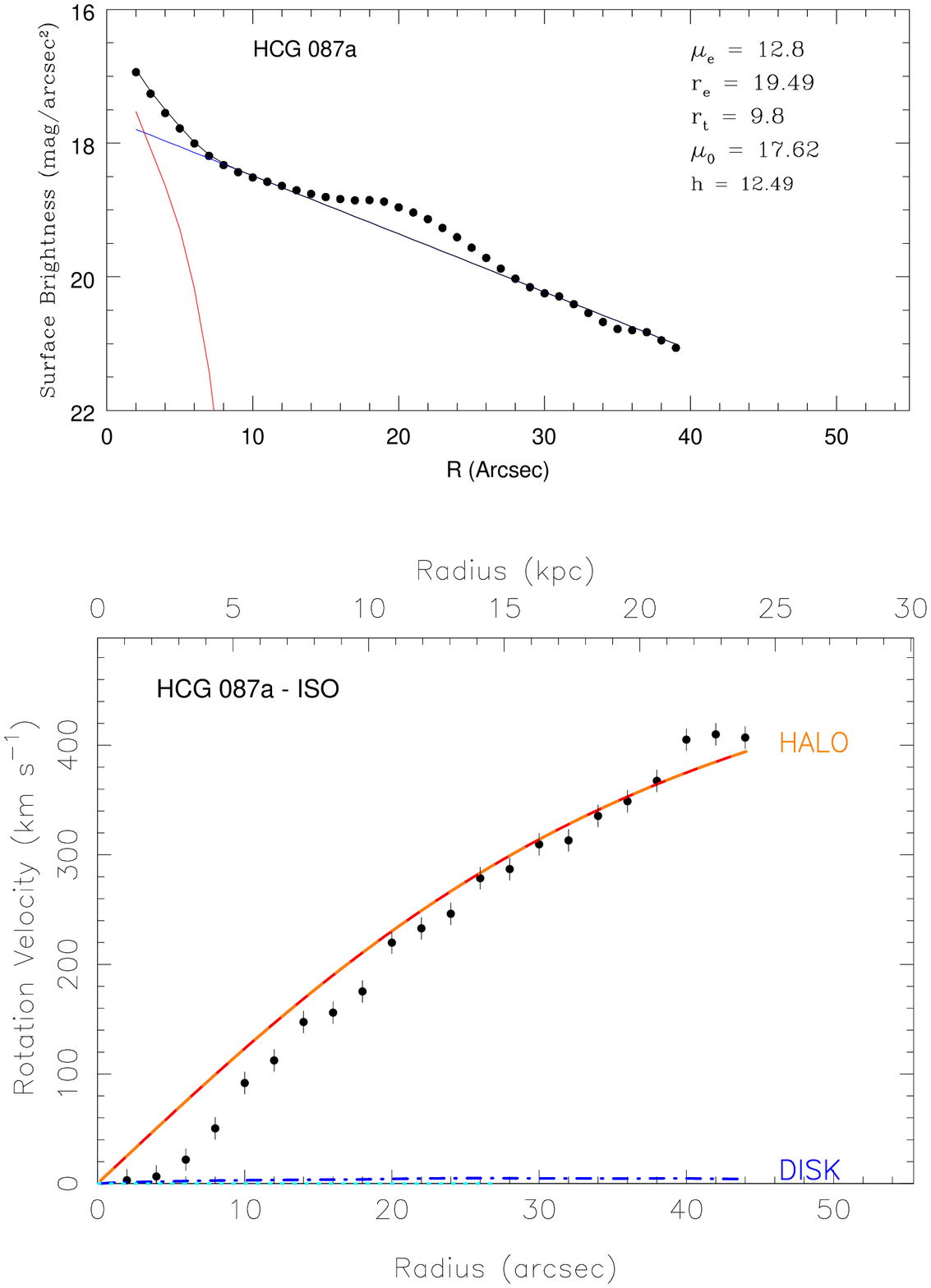}
\caption{\label{Figure10}}
\end{figure*}

\clearpage

\begin{figure*}

\figurenum{11}
\plotone{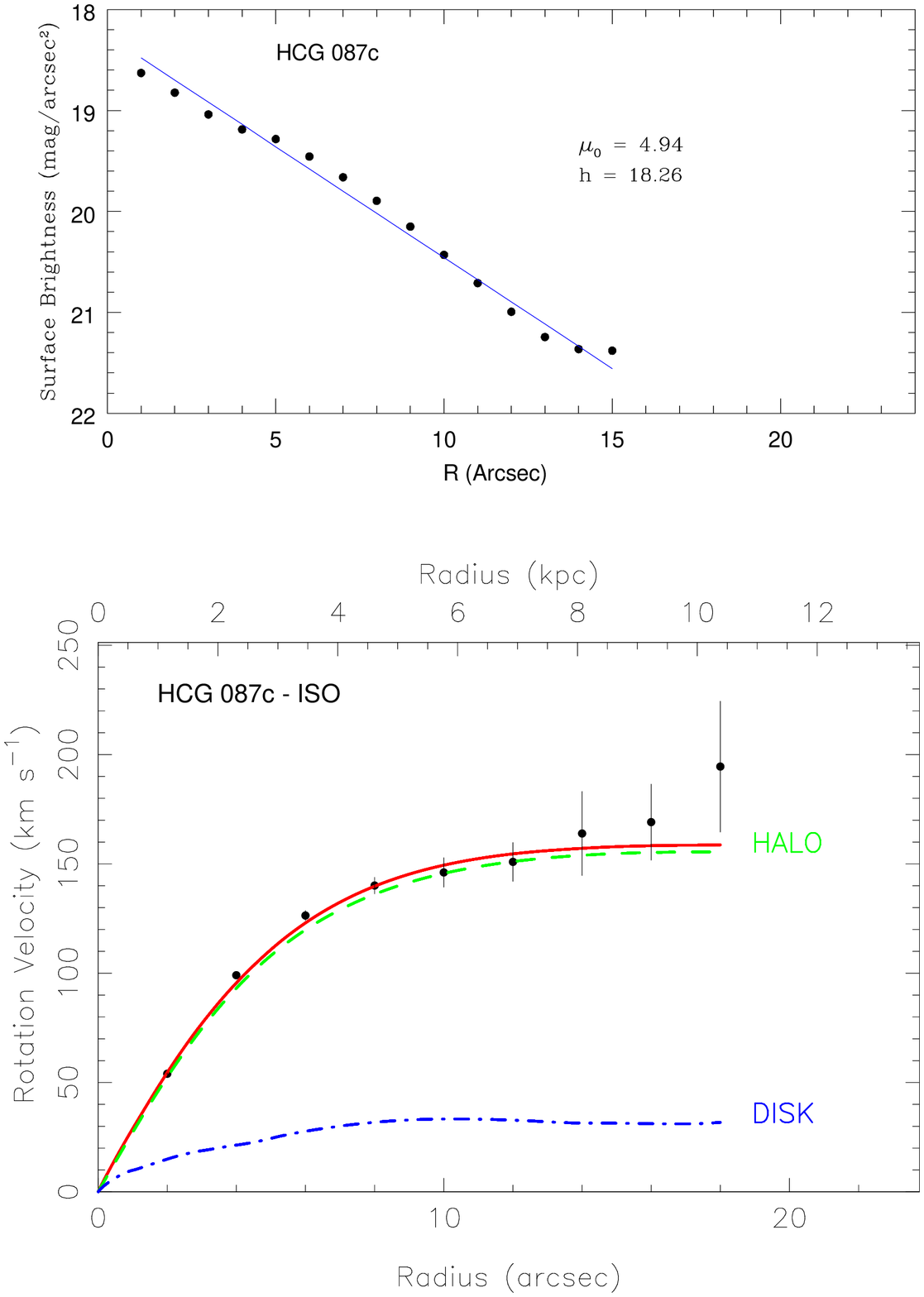}
\caption{\label{Figure11}}
\end{figure*}

\clearpage

\begin{figure*}
\figurenum{12}
\plotone{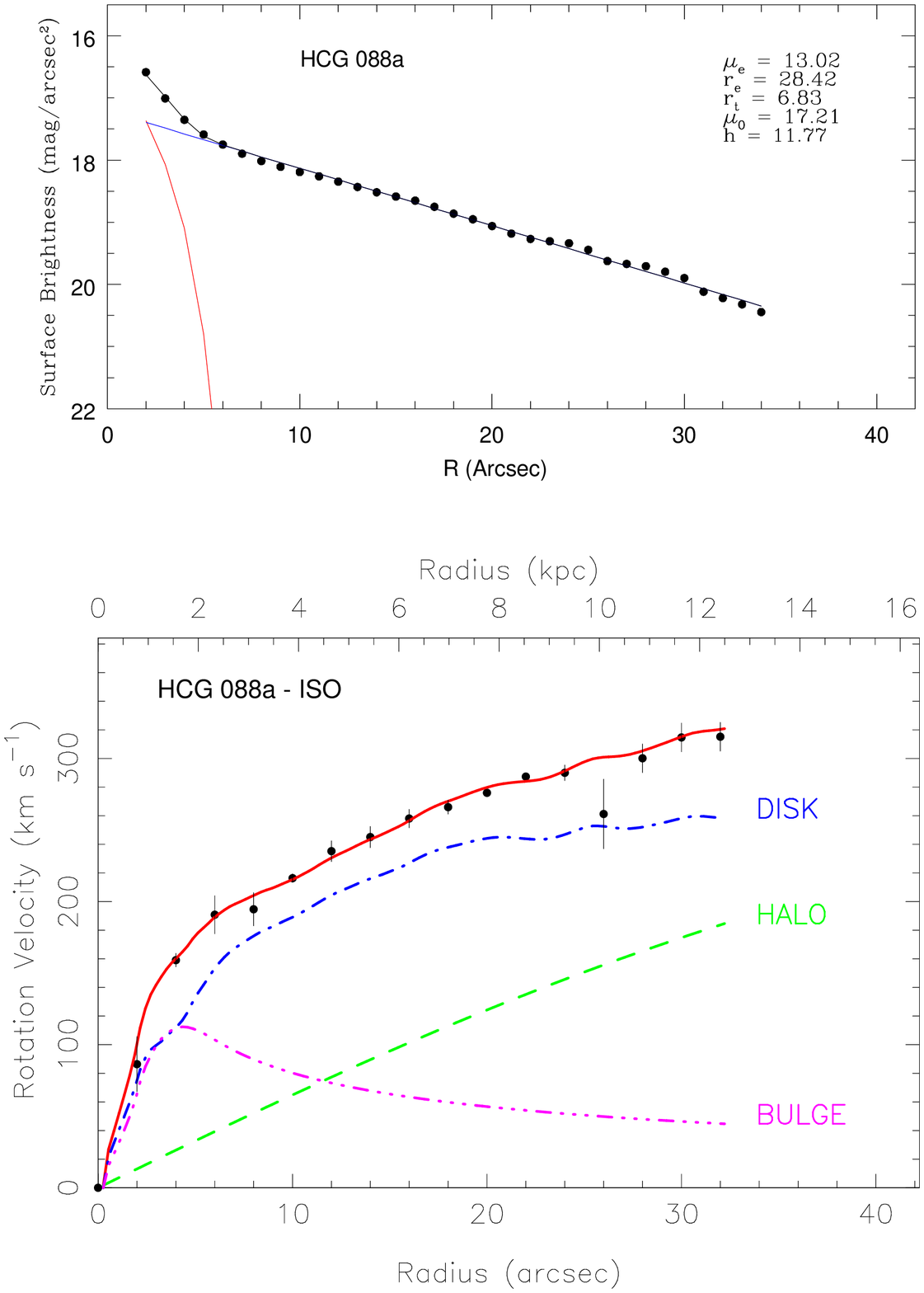}
\caption{\label{Figure12}}
\end{figure*}

\clearpage

\begin{figure*}
\figurenum{13}
\plotone{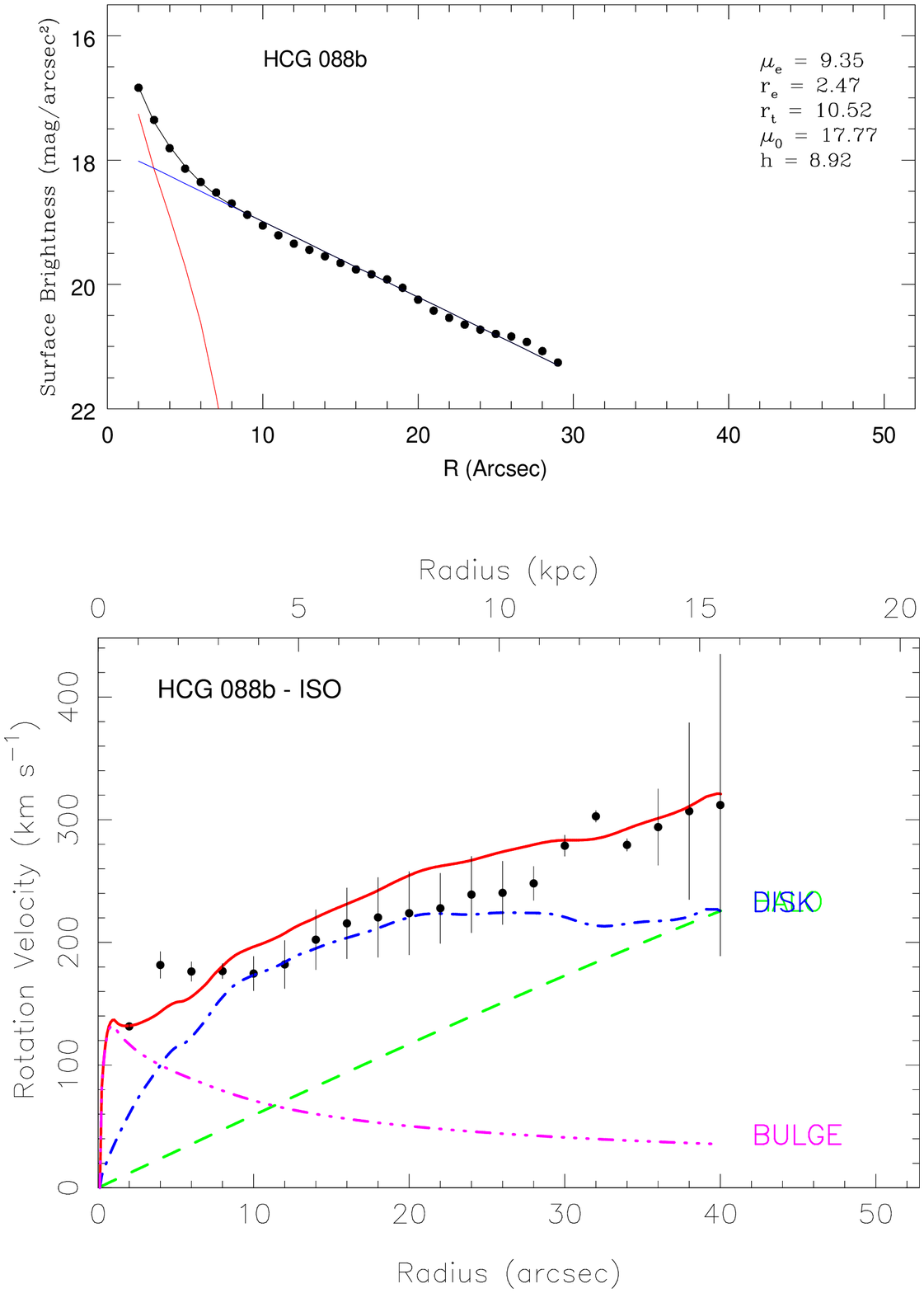}
\caption{\label{Figure13}}
\end{figure*}

\clearpage

\begin{figure*}
\figurenum{14}
\plotone{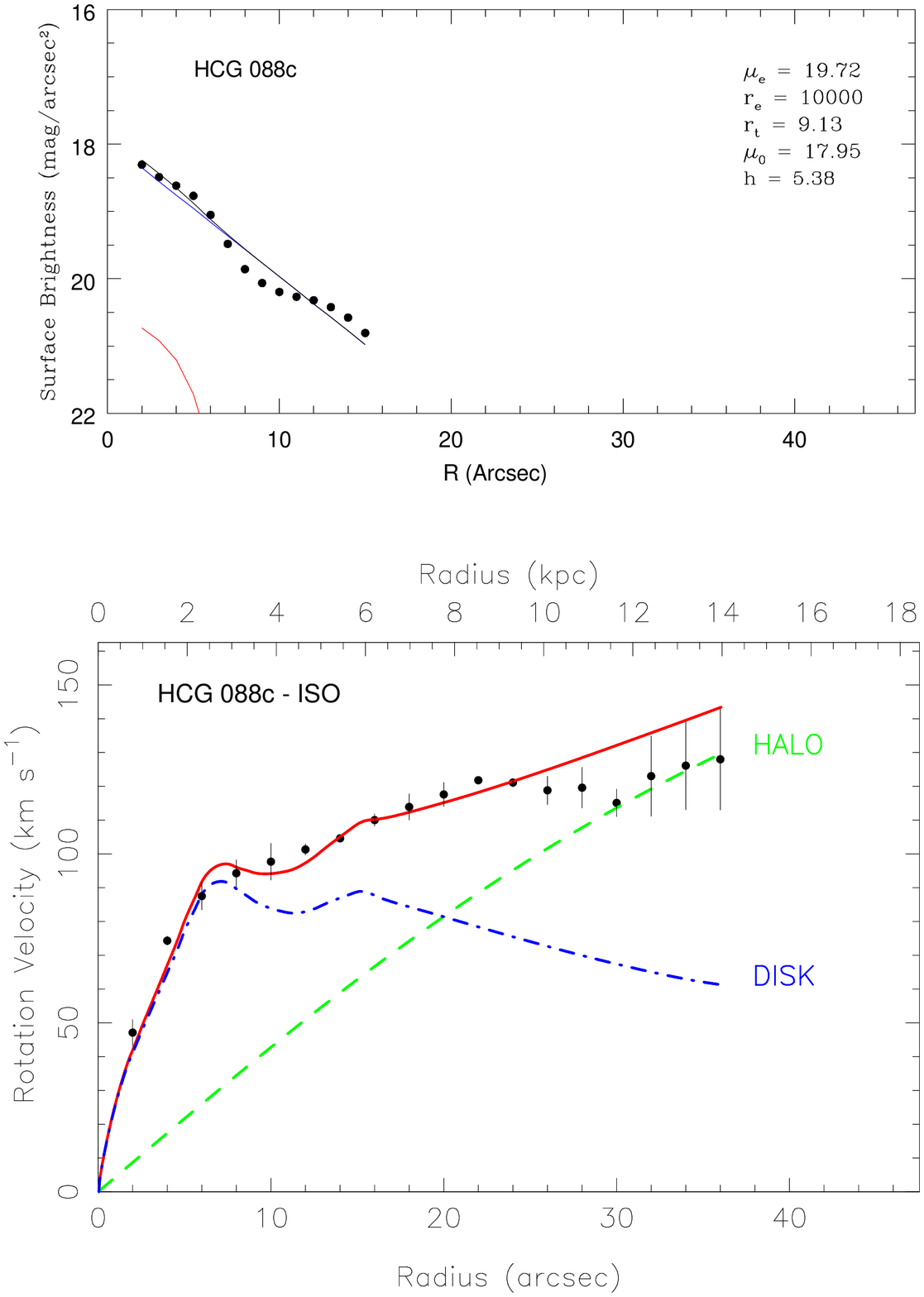}
\caption{\label{Figure14}}
\end{figure*}

\clearpage

\begin{figure*}
\figurenum{15}
\plotone{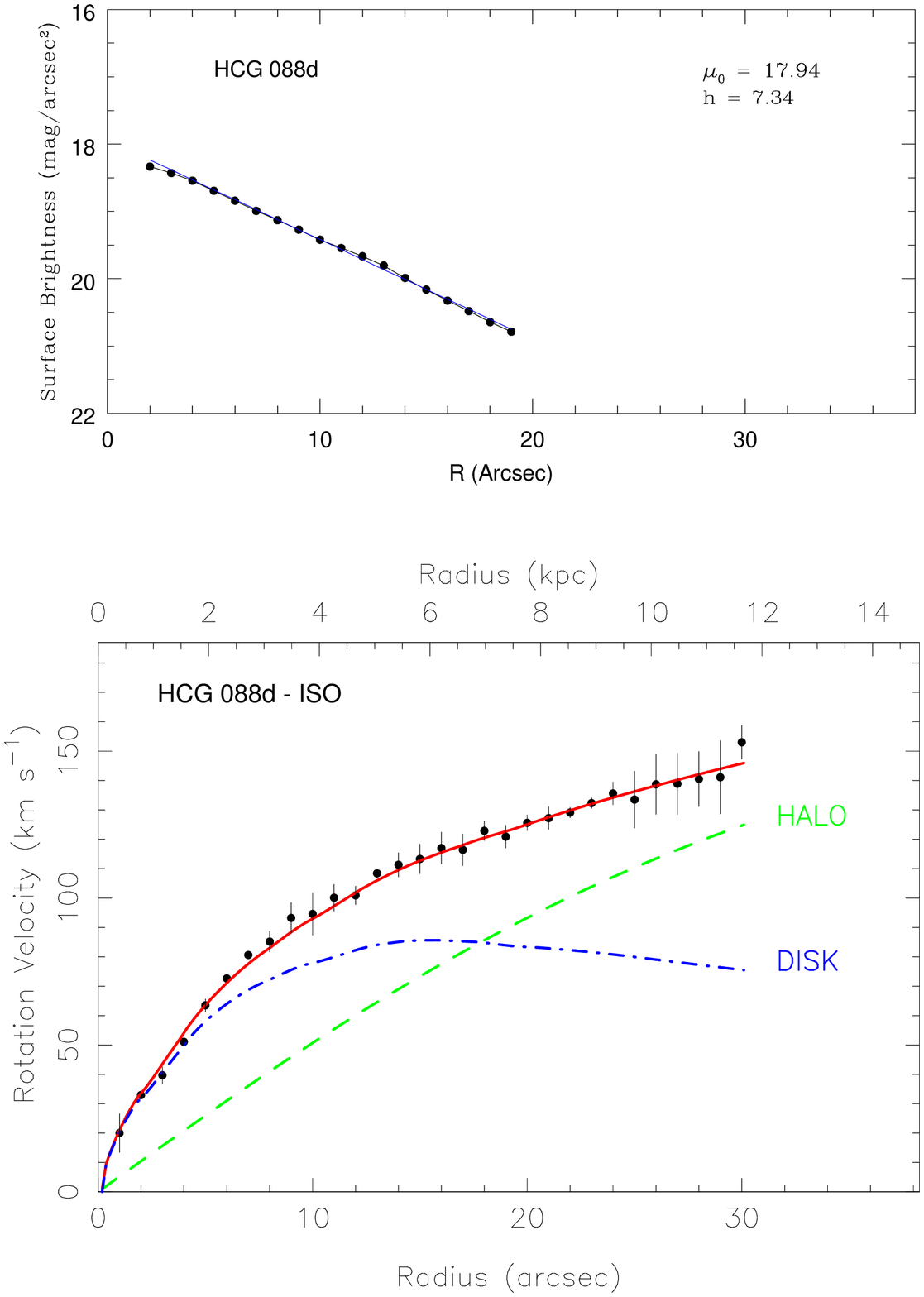}
\caption{\label{Figure15}}
\end{figure*}

\clearpage

\begin{figure*}
\figurenum{16}
\plotone{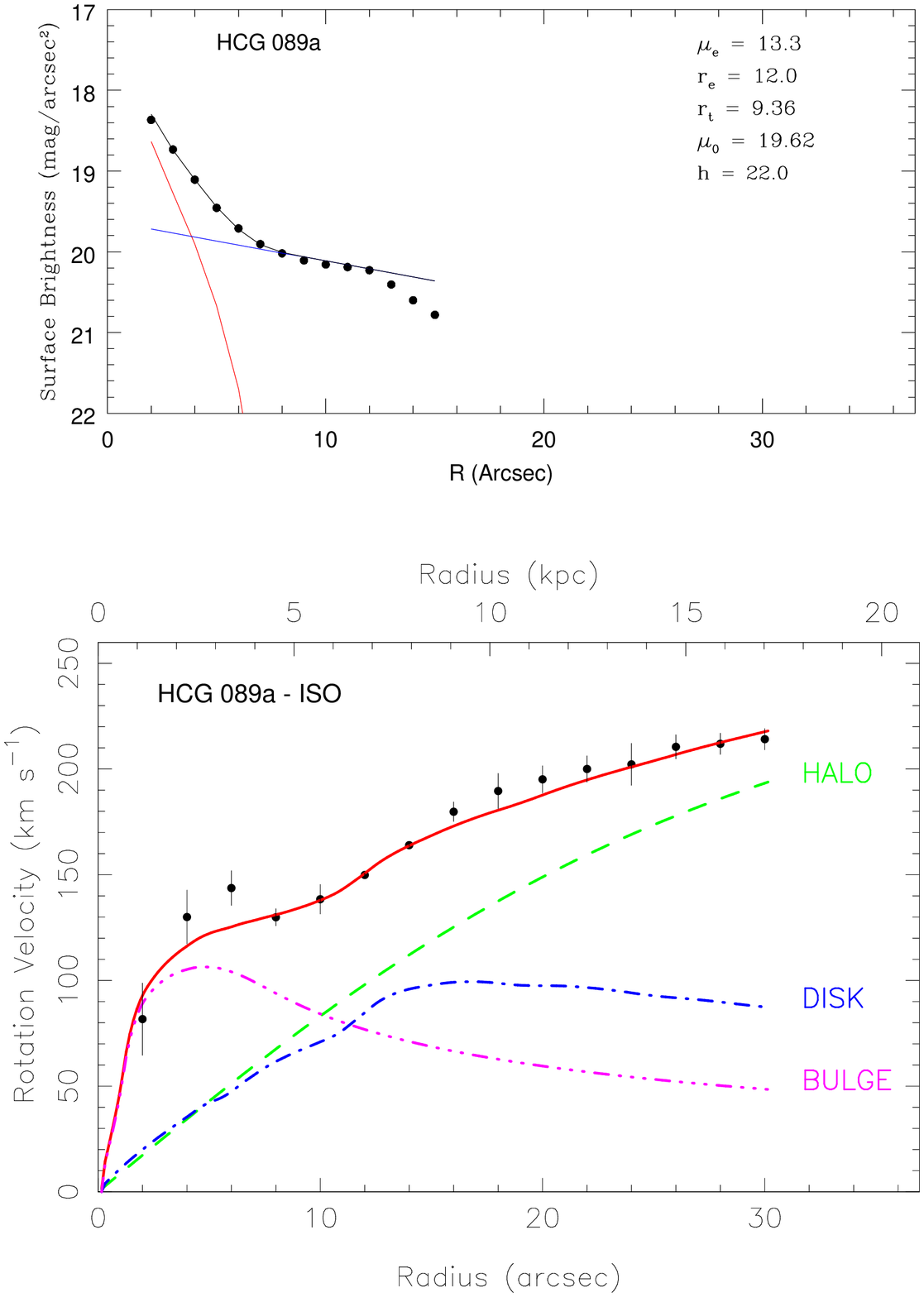}
\caption{\label{Figure16}}
\end{figure*}

\clearpage

\begin{figure*}
\figurenum{17}
\plotone{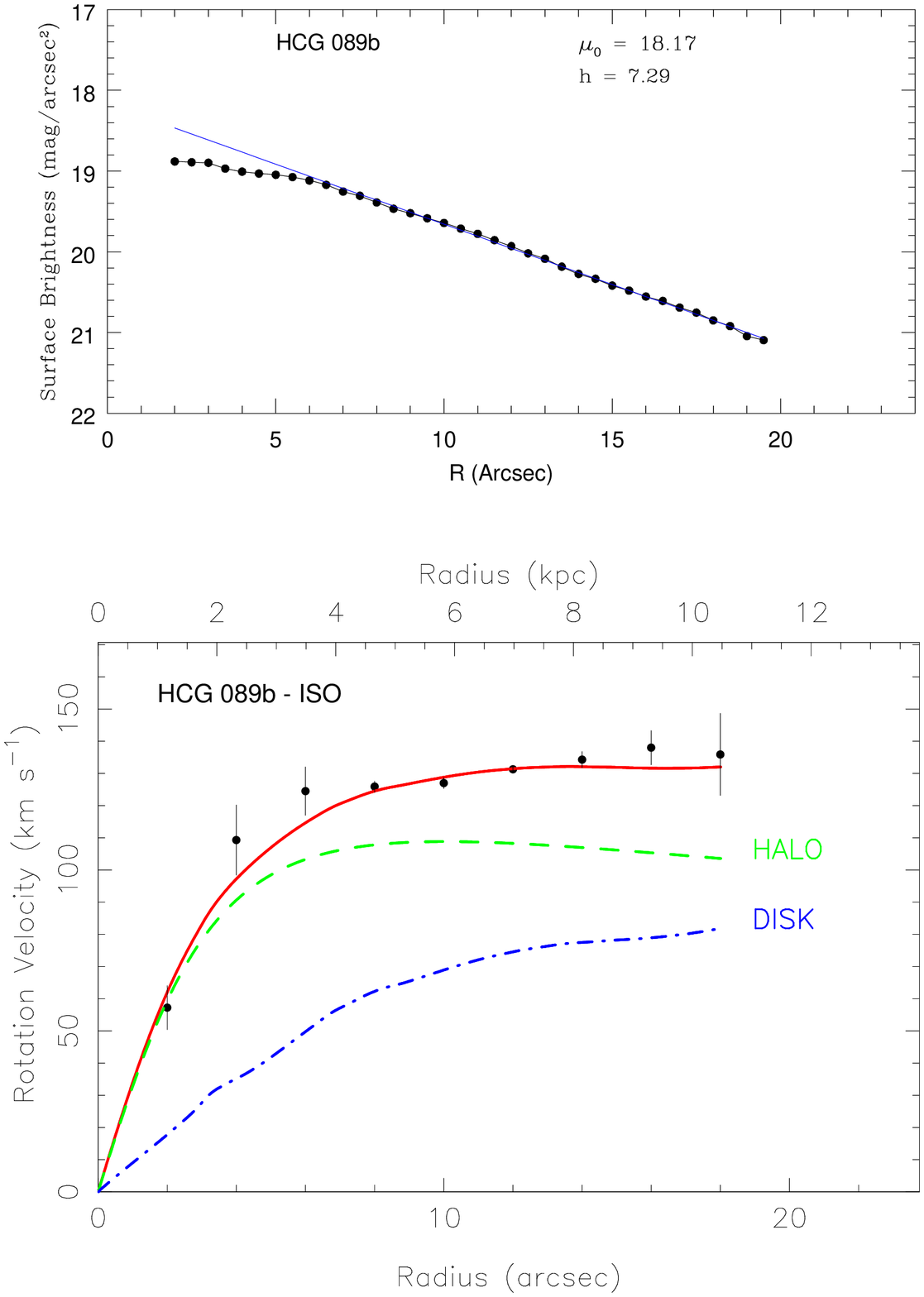}
\caption{\label{Figure17}}
\end{figure*}

\clearpage

\begin{figure*}
\figurenum{18}
\plotone{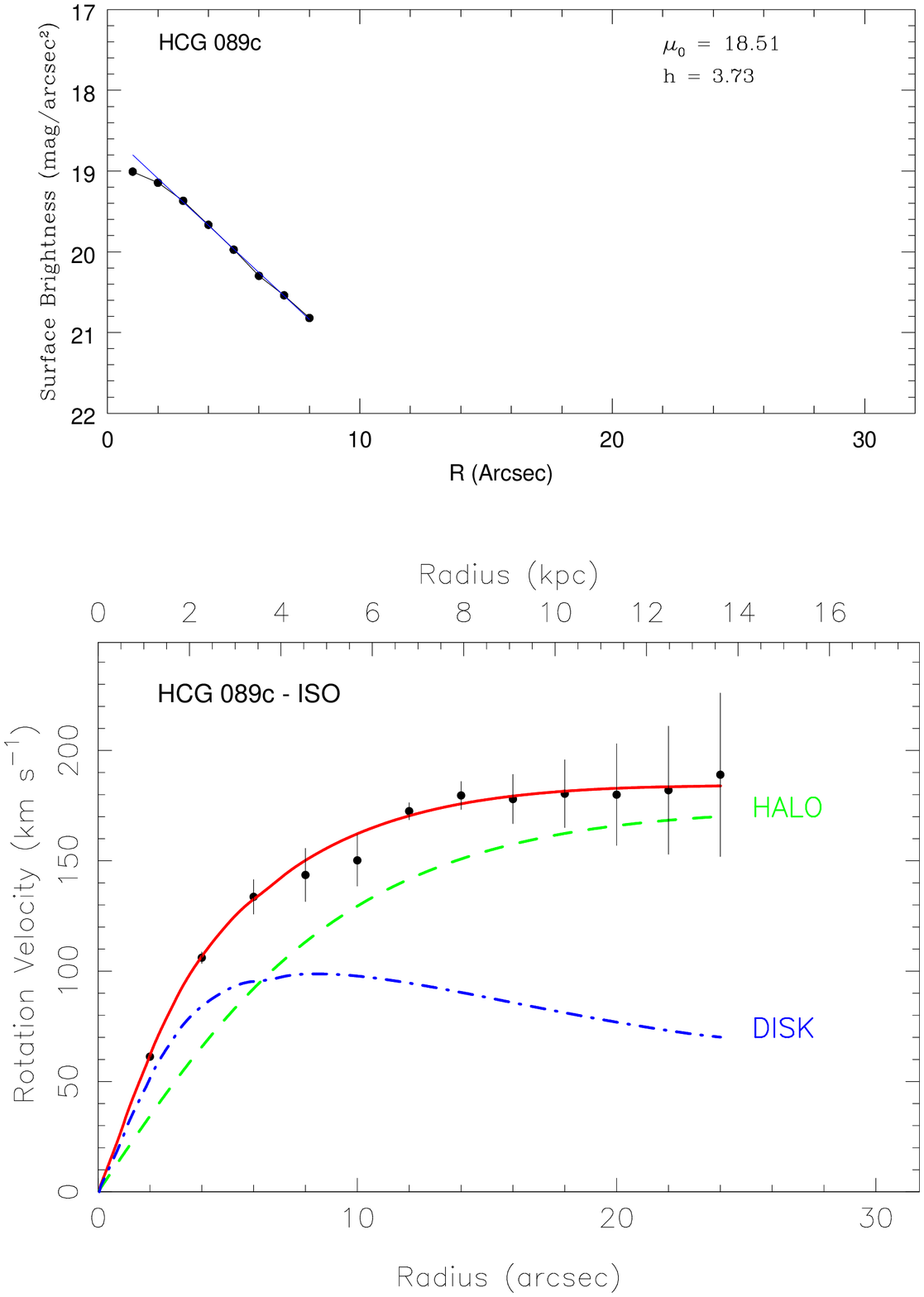}
\caption{\label{Figure18}}
\end{figure*}

\clearpage

\begin{figure*}
\figurenum{19}
\plotone{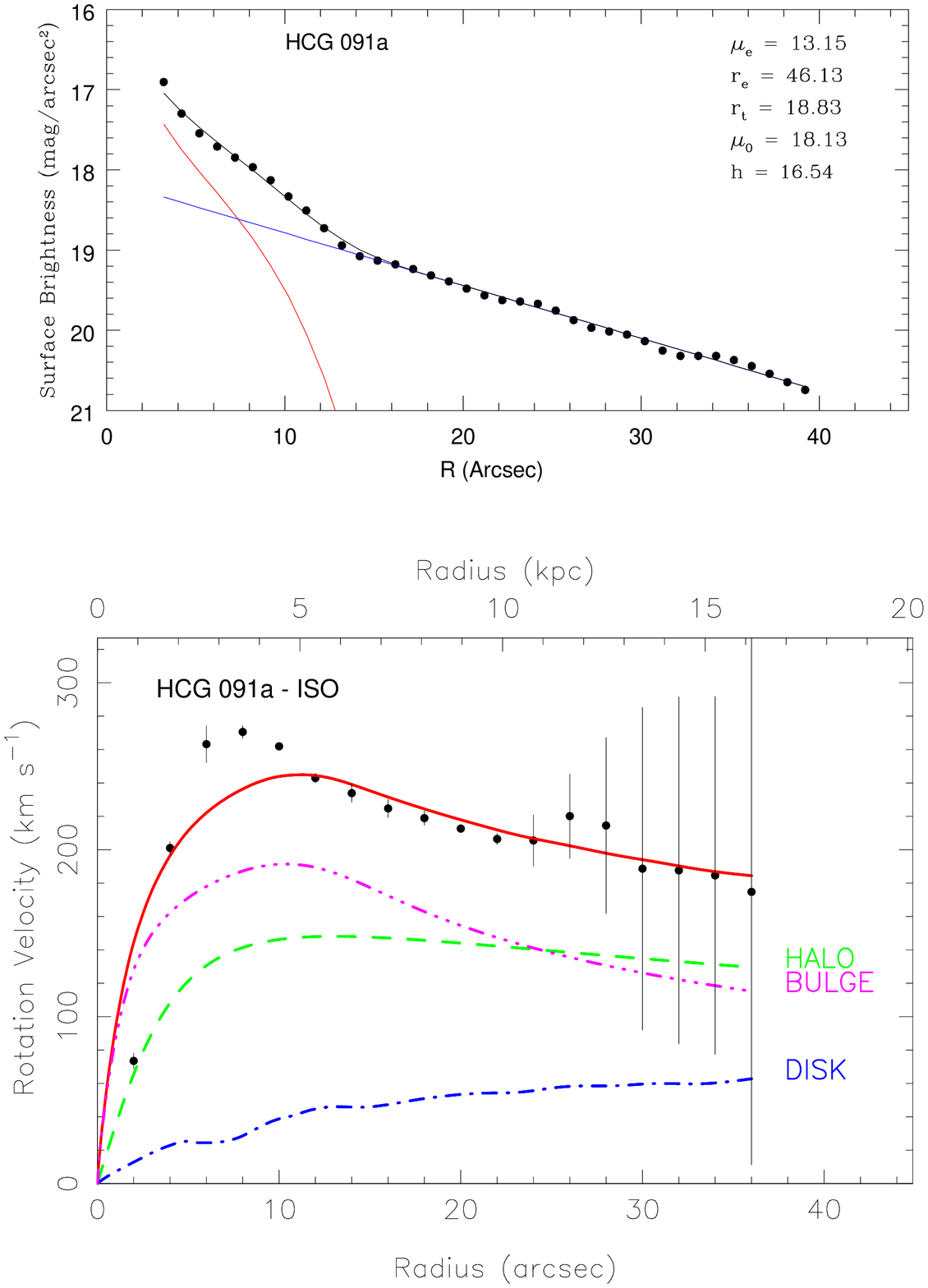}
\caption{\label{Figure19}}
\end{figure*}

\clearpage

\begin{figure*}
\figurenum{20}
\plotone{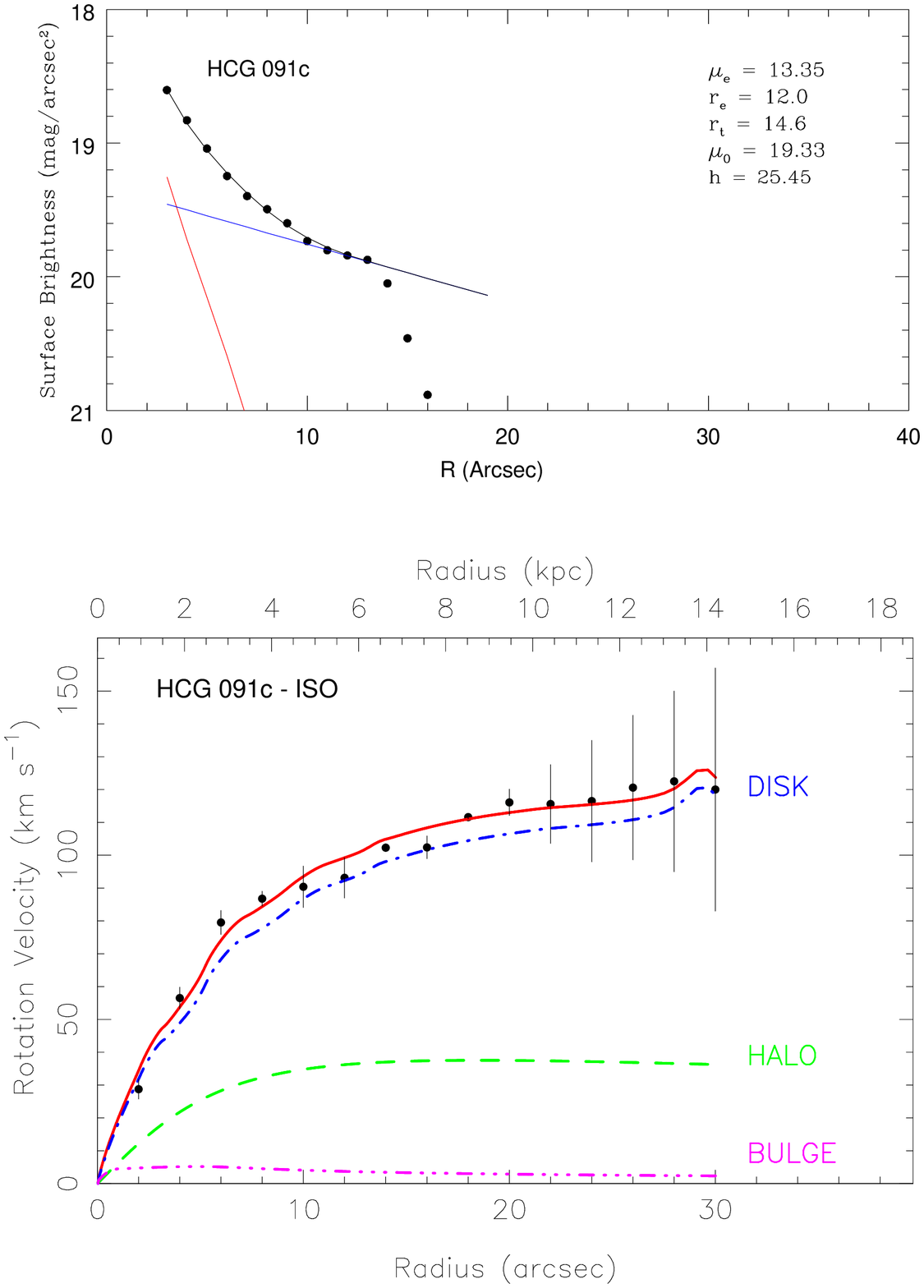}
\caption{\label{Figure20}}
\end{figure*}

\clearpage

\begin{figure*}
\figurenum{21}
\plotone{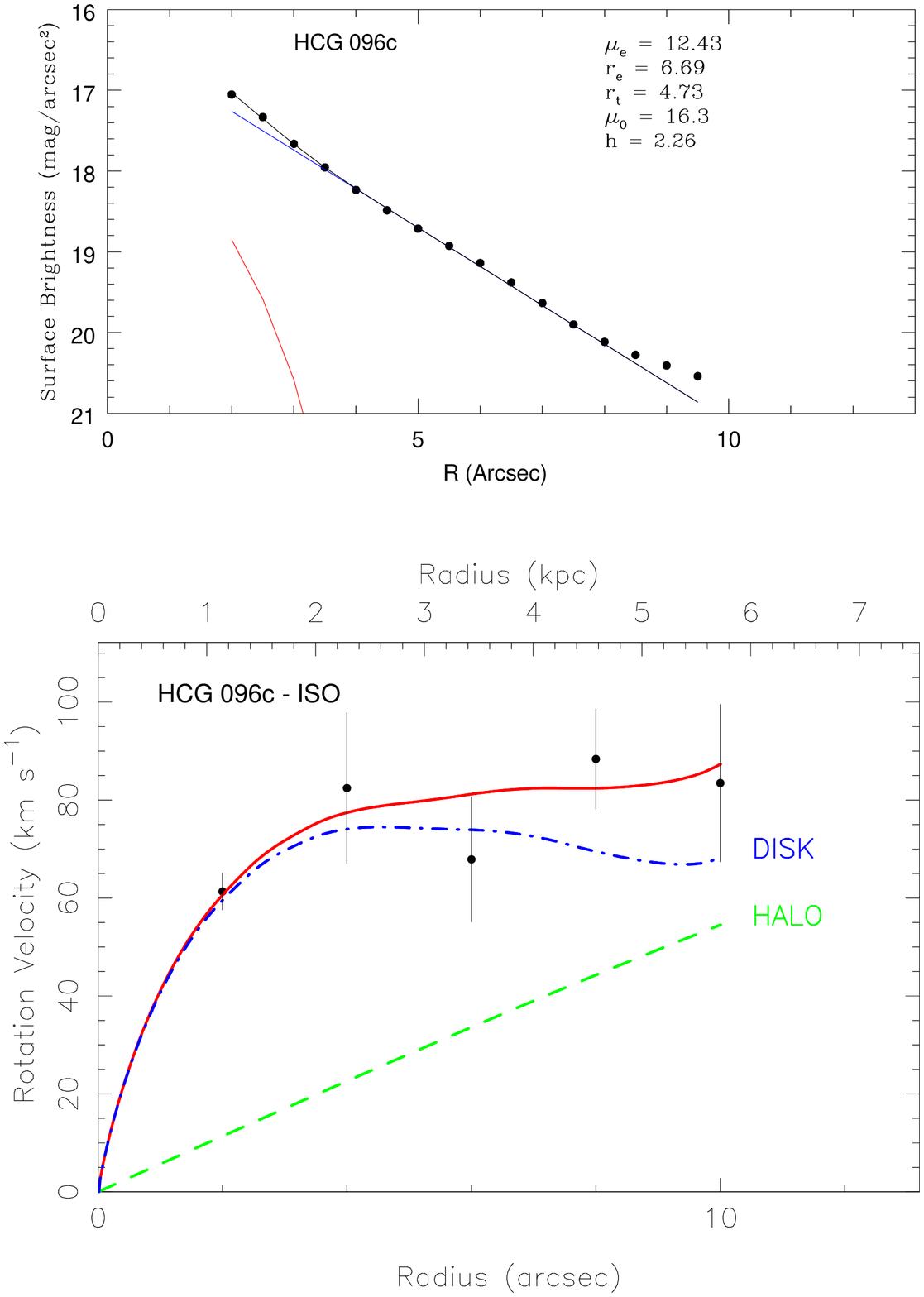}
\caption{\label{Figure21}}
\end{figure*}

\clearpage

%
%

\begin{deluxetable}{lccccccccccc}
\small
\tablenum{1} 
\tablewidth{0pt}

\tablecaption{Photometric and Mass Model parameters for the HCG with the isothermal sphere as the halo J-band}

\tablecolumns{12}

\tablehead{
\colhead{Name} & \colhead{} &\multicolumn{5}{c}{Photometric Parameters}  & \multicolumn{4}{c}{Mass Model Parameters} & \colhead{} \\
\colhead{} & \colhead{} &\colhead{} & \colhead{} & \colhead{} & \colhead{} & \colhead{} & \colhead{} & \colhead{} & \colhead{} & \colhead{} & \colhead{}  \\
\colhead{HCG$^1$} & \colhead{Inclination} &\colhead{$\mu _0$$^2$} & \colhead{$h$$^3$} & \colhead{$\mu_e$$^4$} & \colhead{$r_e$$^5$} & \colhead{$r_t$$^6$} & \colhead{$M/L_{bulge}$$^7$} & \colhead{$M/L_{disk}$$^8$} & \colhead{$r_0$$^9$} & \colhead{$\rho_0$$^{10}$} & \colhead{$\chi^2$$_{11}$} }

\startdata

h007c & $48\pm5$&19.12 &  7.21 & 14.70 &  9.02 &  2.81 &  0.55$ _{ 0.44} ^{ 0.46}$ &  0.05$ _{ 1.41} ^{ 1.49}$ &  6.00$ _{ 4.88} ^{ 5.12}$ &  0.05$ _{ 0.02} ^{ 0.03}$ &  4.29 \\ 
h010d & $68\pm5$ &17.80 &  2.47 &  0.00 &  0.00 &  0.00 &  1.00$ _{ 0.98} ^{ 1.02}$ &  1.60$ _{ 1.85} ^{ 1.95}$ &  1.80$ _{ 1.66} ^{ 1.74}$ & 0.18$ _{ 0.15} ^{ 0.15}$  &  0.39 \\ 
h016a & $43\pm5$  &17.64 &  3.98 &  9.91 &  3.17 &  3.12 &  0.20$ _{ 0.20} ^{ 0.20}$ &  0.40$ _{ 0.73} ^{ 0.77}$ &  1.60$ _{ 1.07} ^{ 1.13}$ & 0.50$ _{ 0.44} ^{ 0.46}$  &  7.61 \\ 
h016c & $60\pm5$ &16.60 &  1.44 & 12.35 &  9.92 &  2.86 &  0.11$ _{ 0.05} ^{ 0.05}$ &  0.03$ _{ 0.10} ^{ 0.10}$ &  6.00$ _{ 4.88} ^{ 5.12}$ &  0.13$ _{ 0.16} ^{ 0.16}$ &  3.00 \\ 
h019a & $53\pm10$ & 0.00 &  1.00 &  9.95 &  1.88 & 25.06 &  0.00$ _{ 0.00} ^{ 0.00}$ &  0.10$ _{ 0.10} ^{ 0.10}$ &  9.00$ _{ 8.78} ^{ 9.22}$ &  0.02$ _{ 0.03} ^{ 0.03}$ & 39.53 \\ 
h019b & $60\pm10$ &18.75 &  3.12 & 11.90 &  0.84 &  2.71 &  0.14$ _{ 0.02} ^{ 0.03}$ &  0.05$ _{ 0.20} ^{ 0.22}$ &  8.00$ _{17.55} ^{18.45}$ & 0.01$ _{ 0.01} ^{ 0.01}$  &  3.13 \\ 
h087a & $85\pm10$ &17.62 &  6.78 & 12.80 & 10.58 &  5.32 &  0.00$ _{ 0.00} ^{ 0.00}$ &  0.00$ _{ 0.38} ^{ 0.40}$ & 23.00$ _{ 4.88} ^{ 5.12}$ & 0.03$ _{ 0.01} ^{ 0.01}$  &  8.22 \\ 
h087c & $50\pm3$ &18.26 &  2.85 &  0.00 &  0.00 &  0.00 &  0.10$ _{ 0.10} ^{ 0.10}$ &  0.10$ _{ 0.20} ^{ 0.20}$ &  3.50$ _{ 3.90} ^{ 4.10}$ &  0.13$ _{ 0.11} ^{ 0.11}$ &  2.07 \\ 
h088a & $65\pm5$ &17.21 &  4.56 & 13.01 & 11.02 &  2.65 &  1.05$ _{ 1.02} ^{ 1.08}$ &  1.75$ _{ 1.71} ^{ 1.79}$ & 21.00$ _{35.10} ^{36.90}$ & 0.02$ _{ 0.01} ^{ 0.01}$  &  0.90 \\ 
h088b & $55\pm3$ &18.13 &  4.56 &  9.10 &  1.12 &  4.27 &  0.20$ _{ 0.20} ^{ 0.20}$ &  1.50$ _{ 1.75} ^{ 1.84}$ & 45.00$ _{74.10} ^{77.90}$ & 0.01$ _{ 0.01} ^{ 0.01}$  &  3.64 \\ 
h088c & $42\pm2$ &17.95 &  2.09 & 19.72 & 3882.00 &  3.54 &  0.30$ _{ 0.29} ^{ 0.31}$ &  0.45$ _{ 0.48} ^{ 0.50}$ & 19.50$ _{22.91} ^{24.09}$ & 0.01$ _{ 0.01} ^{ 0.01}$  &  5.89 \\
h088d & $70\pm2$ &17.94 &  2.85 &  0.00 &  0.00 &  0.00 &  0.10$ _{ 0.10} ^{ 0.10}$ &  0.97$ _{ 0.98} ^{ 1.02}$ & 14.50$ _{14.14} ^{14.86}$ & 0.01$ _{ 0.01} ^{ 0.01}$  &  1.20 \\
h089a & $45\pm5$ &19.62 & 12.48 & 13.30 &  6.81 &  5.31 &  1.20$ _{ 1.07} ^{ 1.13}$ &  0.80$ _{ 0.88} ^{ 0.92}$ & 18.00$ _{19.50} ^{20.50}$ & 0.01$ _{ 0.01} ^{ 0.01}$  &  1.28 \\ 
h089b & $49\pm3$ &18.17 &  4.24 &  0.00 &  0.00 &  0.00 &  1.00$ _{ 0.98} ^{ 1.02}$ &  0.70$ _{ 0.98} ^{ 1.02}$ &  2.00$ _{ 1.46} ^{ 1.54}$ & 0.19$ _{ 0.28} ^{ 0.29}$  &  1.28 \\
h089c & $46\pm3$ &18.51 &  2.12 &  0.00 &  0.00 &  0.00 &  1.00$ _{ 0.98} ^{ 1.02}$ &  1.10$ _{ 1.17} ^{ 1.23}$ &  6.00$ _{ 7.80} ^{ 8.20}$ & 0.05$ _{ 0.04} ^{ 0.04}$  &  0.28 \\
h091a & $50\pm3$ &18.13 &  7.42 & 13.15 & 20.69 &  8.44 &  1.00$ _{ 0.98} ^{ 1.02}$ &  0.10$ _{ 0.20} ^{ 0.20}$ &  2.00$ _{ 1.95} ^{ 2.05}$ & 0.35$ _{ 0.31} ^{ 0.33}$  & 33.14 \\
h091c & $40\pm5$ &18.76 &  5.08 & 14.80 & 12.78 &  4.26 &  0.00$ _{ 0.01} ^{ 0.01}$ &  0.70$ _{ 0.76} ^{ 0.80}$ &  3.00$ _{ 1.95} ^{ 2.05}$ & 0.01$ _{ 0.01} ^{ 0.01}$  &  1.66 \\
h096c & $57\pm5$ &16.30 &  1.29 & 12.43 &  3.83 &  2.50 &  0.05$ _{ 0.05} ^{ 0.05}$ &  0.16$ _{ 0.16} ^{ 0.16}$ & 18.00$ _{35.10} ^{36.90}$ &  0.01$ _{ 0.01} ^{ 0.01}$ &  0.81 \\

\enddata
\bigskip

\tablecomments{
$^1$ Hickson Group,
$^2$ Inclination of the disk from velocity fields,
$^3$ Disk central surface brightness in $mag/arcsec^2$, 
$^4$ Scale length in kpc,
$^5$ Bulge central surface brightness in $mag/arcsec^2$,  
$^6$ Characteristic radius in kpc, 
$^7$ Generalized characteristic radius kpc,
$^8$ Mass / Luminosity ratio for the bulge in M$_{\odot}$/L$_{\odot}$,
$^9$ Mass / Luminosity ratio for the disk in M$_{\odot}$/L$_{\odot}$,
$^{10}$ Characteristic radius for the Dark Halo in kpc,
$^{11}$ Central density for the dark halo in M$_{\odot}$~pc$^{-3}$, 
$^{12}$ Reduced $\chi^2$
}

\end{deluxetable}

\clearpage

%
%

\begin{deluxetable}{lcccccccccc}
\small
\tablenum{2} \tablewidth{0pt}

\tablecaption{Halo and M/L parameters for the HCG with Maximum Disk and NFW models}

\tablecolumns{11}

\tablehead{
\colhead{Name} &  \multicolumn{5}{c}{Halo \& M/L Parameters}  & \multicolumn{5}{c}{Halo \& M/L Parameters} \\
\colhead{} &  \multicolumn{5}{c}{MDM Model}  & \multicolumn{5}{c}{NFW Model} \\
\colhead{} & \colhead{} & \colhead{} & \colhead{} & \colhead{} & \colhead{} & \colhead{} & \colhead{} & \colhead{} & \colhead{}  \\
\colhead{HCG$^1$} & \colhead{$M/L_{bulge}$$^2$} & \colhead{$M/L_{disk}$$^3$} & \colhead{$r_0$$^4$} & \colhead{$\rho_0$$^{5}$} & \colhead{$\chi^2$} & \colhead{$M/L_{bulge}$$^6$} & \colhead{$M/L_{disk}$$^7$} & \colhead{$r_0$$^8$} & \colhead{$\rho_0$$^{9}$} & \colhead{$\chi^2$$_{10}$} }

\startdata

h007c &  0.45 &  1.45 &  5.00 &  0.03 &  5.13 &  0.00$ _{ 0.00} ^{ 0.01}$ &  1.95$ _{ 1.90} ^{ 2.05}$ &  6.00$ _{ 5.85} ^{ 6.15}$ &  0.01$ _{ 0.01} ^{ 0.01}$ &  5.50 \\ 
h010d &  1.00 &  1.90 &  1.70 &  0.15 &  0.43 &  1.00$ _{ 0.98} ^{ 1.02}$ &  2.70$ _{ 2.63} ^{ 2.77}$ &  2.50$ _{ 2.10} ^{ 2.80}$ & 0.03$ _{ 0.02} ^{ 0.03}$   &  1.42 \\ 
h016a &  0.20 &  0.75 &  1.10 &  0.45 &  9.04 &  0.15$ _{ 0.15} ^{ 0.15}$ &  0.35$ _{ 0.34} ^{ 0.36}$ &  1.70$ _{ 1.66} ^{ 1.74}$ & 0.75$ _{ 0.73} ^{ 0.77}$    &  7.54 \\ 
h016c &  0.05 &  0.10 &  5.00 &  0.16 &  3.52 &  0.00$ _{ 0.00} ^{ 0.05}$ &  0.01$ _{ 0.01} ^{ 0.04}$ & 35.00$ _{34.12} ^{35.88}$ &  0.01$ _{ 0.01} ^{ 0.01}$ & 14.27 \\ 
h019a &  0.00 &  0.10 &  9.00 &  0.03 & 45.26 &  0.00$ _{ 0.00} ^{ 0.00}$ &  0.10$ _{ 0.10} ^{ 0.10}$ & 12.00$ _{11.70} ^{12.30}$ &  0.00$ _{ 0.00} ^{ 0.00}$   & 189.79 \\ 
h019b &  0.03 &  0.21 & 18.00 &  0.01 &  3.65 &  0.00$ _{ 0.00} ^{ 0.05}$ &  0.00$ _{ 0.00} ^{ 0.04}$ & 32.00$ _{29.00} ^{35.00}$ & 0.00$ _{ 0.00} ^{ 0.00}$  &  7.53 \\ 
h087a &  0.00 &  0.39 &  5.00 &  0.01 &  1.19 &  0.00$ _{ 0.00} ^{ 0.00}$ &  0.01$ _{ 0.01} ^{ 0.15}$ & 54.00$ _{52.00} ^{55.00}$ & 0.01$ _{ 0.01} ^{ 0.01}$  & 59.36 \\ 
h087c &  0.10 &  0.20 &  4.00 &  0.11 &  2.47 &  0.10$ _{ 0.10} ^{ 0.10}$ &  1.40$ _{ 1.37} ^{ 1.43}$ &  1.00$ _{ 0.98} ^{ 1.02}$ & 0.01$ _{ 0.01} ^{ 0.01}$  & 63.94 \\ 
h088a &  1.05 &  1.75 & 36.00 &  0.01 &  1.03 &  0.80$ _{ 0.55} ^{ 1.05}$ &  1.00$ _{ 0.98} ^{ 1.02}$ & 130.00$ _{126.75} ^{133.25}$ & 0.00$ _{ 0.00} ^{ 0.00}$  &  0.72 \\ 
h088b &  0.20 &  1.80 & 76.00 &  0.01 &  4.33 &  0.10$ _{ 0.10} ^{ 0.10}$ &  0.50$ _{ 0.49} ^{ 0.70}$ & 16.00$ _{15.60} ^{16.40}$ & 0.03$ _{ 0.02} ^{ 0.03}$  &  4.54 \\ 
h088c &  0.30 &  0.49 & 23.50 &  0.01 &  6.91 &  0.30$ _{ 0.29} ^{ 0.31}$ &  0.03$ _{ 0.01} ^{ 0.04}$ & 13.00$ _{12.68} ^{13.32}$ & 0.01$ _{ 0.01} ^{ 0.01}$  &  5.33 \\ 
h088d &  0.10 &  1.00 & 14.50 &  0.01 &  1.44 &  0.10$ _{ 0.10} ^{ 0.10}$ &  1.39$ _{ 1.34} ^{ 1.45}$ & 19.95$ _{17.95} ^{20.45}$ &  0.00$ _{ 0.00} ^{ 0.00}$ & 61.71 \\ 
h089a &  1.10 &  0.90 & 20.00 &  0.01 &  1.43 &  0.15$ _{ 0.00} ^{ 0.25}$ &  0.11$ _{ 0.07} ^{ 0.15}$ & 67.00$ _{65.33} ^{68.67}$ & 0.00$ _{ 0.00} ^{ 0.00}$  &  3.76 \\ 
h089b &  1.00 &  1.00 &  1.50 &  0.28 &  1.51 &  1.00$ _{ 0.98} ^{ 1.02}$ &  0.80$ _{ 0.78} ^{ 0.82}$ &  3.00$ _{ 2.93} ^{ 3.07}$ & 0.10$ _{ 0.10} ^{ 0.11}$  &  3.22 \\ 
h089c &  1.00 &  1.20 &  8.00 &  0.04 &  0.33 &  1.00$ _{ 0.98} ^{ 1.02}$ &  0.01$ _{ 0.01} ^{ 0.11}$ & 97.00$ _{92.00} ^{101.00}$ & 0.00$ _{ 0.00} ^{ 0.00}$  &  9.88 \\ 
h091a &  1.00 &  0.20 &  2.00 &  0.32 & 37.54 &  1.60$ _{ 1.56} ^{ 1.64}$ &  0.01$ _{ 0.01} ^{ 0.01}$ & 10.00$ _{ 9.00} ^{10.25}$ &  0.01$ _{ 0.01} ^{ 0.01}$ & 36.49 \\ 
h091c &  0.01 &  0.78 &  2.00 &  0.01 &  1.95 &  0.00$ _{ 0.00} ^{ 0.09}$ &  0.79$ _{ 0.77} ^{ 0.81}$ &  1.00$ _{ 0.98} ^{ 5.00}$ &   0.00$ _{ 0.00} ^{ 0.01}$  &  1.74 \\ 
h096c &  0.05 &  0.16 & 36.00 &  0.01 &  0.98 &  0.05$ _{ 0.05} ^{ 0.05}$ &  0.11$ _{ 0.11} ^{ 0.11}$ &  9.00$ _{ 7.00} ^{11.00}$ &   0.01$ _{ 0.00} ^{ 0.01}$  &  0.82 \\ 

\enddata
\bigskip

\tablecomments{
$^1$ Hickson Group, 
$^2$ Mass / Luminosity ratio for the disk in M$_{\odot}$/L$_{\odot}$ using MDM model, 
$^3$ Mass / Luminosity ratio for the bulge in M$_{\odot}$/L$_{\odot}$ using MDM model, 
$^4$ Characteristic radius for the Dark Halo in kpc using MDM model, 
$^5$ Central density for the dark halo in M$_{\odot}$~pc$^{-3}$ using MDM model, 
$^6$ Mass / Luminosity ratio for the bulge in M$_{\odot}$/L$_{\odot}$ using NFW model, 
$^7$ Mass / Luminosity ratio for the disk in M$_{\odot}$/L$_{\odot}$ using NFW model, 
$^8$ Characteristic radius for the Dark Halo in kpc using NFW model,
$^9$ Central density for the dark halo in M$_{\odot}$~pc$^{-3}$ using NFW model,
$^{10}$ Reduced $\chi^2$
}

\end{deluxetable}

\clearpage

%
%

\begin{deluxetable}{lccccccc}
\tablenum{3} \tablewidth{0pt}

\tablecaption{Table summarising the coefficients of the linear regression: $\log(\rho_0) = A log(r_0) - B$.}

\tablecolumns{8}

\tablehead{
\colhead{Coef/Corr} &  \multicolumn{3}{c}{ISO Model}  & \multicolumn{3}{c}{NFW Model} & \colhead{MDM Model} \\
\colhead{} &  \colhead{HCG} &  \colhead{Field} & \colhead{Cluster} &  \colhead{HCG} &  \colhead{Field} & \colhead{Cluster} & \colhead{}  \\
\colhead{} &  \colhead{Galaxies} &  \colhead{Galaxies} & \colhead{Galaxies} &  \colhead{Galaxies} &  \colhead{Galaxies} & \colhead{Galaxies} & \colhead{} } 

\startdata
   A                      &     $-1.0\pm 0.22$     &      $-1.03\pm 0.20$      &     $-1.62\pm 0.12$   & $-0.53\pm0.24 $     & $-1.15\pm 0.30 $    &   $-1.72\pm 0.09$   &  \\
   B                      &     $-0.47\pm 0.21$   &      $-0.68\pm 0.67$      &     $-0.28\pm 0.05$   & $-1.50\pm0.3 $       &  $-0.76\pm 0.20$    &   $-0.20\pm 0.12$   &  \\
   Corr                 &      $-0.80\pm 0.21$   &      $-0.83\pm 0.17 $     &     $-0.92\pm 0.07$   &  $-0.48\pm0.46 $   &  $-0.48\pm 0.37$    &    $-0.98\pm 0.17$  &  $ -0.70 \pm 0.26$ \\

\enddata

\end{deluxetable}

\clearpage

%
%

\begin{deluxetable}{lcccccccccc}
\small

\tablenum{4} \tablewidth{0pt}

\tablecaption{Mass table for HCG in J-band for ISO, MDM and NFW models}

\tablecolumns{11}

\tablehead{
\colhead{} &  \colhead {} & \multicolumn{3}{c}{ISO Model}  & \multicolumn{3}{c}{MDM Model} &\multicolumn{3}{c}{NFW Model} \\
\colhead{} &  \colhead {} & \multicolumn{3}{c}{}  & \multicolumn{3}{c}{} &\multicolumn{3}{c}{} \\
\colhead{HCG} & \colhead{Distance} & \colhead{Disk} & \colhead{Halo} & \colhead{M$_{Halo}$} & \colhead{Disk} & \colhead{Halo} & \colhead{M$_{Halo}$} & \colhead{Disk} & \colhead{Halo} & \colhead{M$_{Halo}$}\\
\colhead{} & \colhead{} & \colhead{Mass} & \colhead{Mass} & \colhead{/} & \colhead{Mass} & \colhead{Mass} & \colhead{/} & \colhead{Mass} & \colhead{Mass} & \colhead{/}  \\
\colhead{} & \colhead{Mpc} & \colhead{10$^{10}$ M$_{\odot}$} & \colhead{10$^{10}$ M$_{\odot}$} & \colhead{M$_{Total}$}  & \colhead{10$^{10}$ M$_{\odot}$} & \colhead{10$^{10}$ M$_{\odot}$} & \colhead{M$_{Total}$} & \colhead{10$^{10}$ M$_{\odot}$} & \colhead{10$^{10}$ M$_{\odot}$} & \colhead{M$_{Total}$} }

\startdata

h007c & 58.00 & 0.18 & 9.5 &  0.98 & 5.1 & 3.3 &  0.39 & 6.8 & 1.40 &  0.17 \\
h010d & 62.00 & 1.5 & 1.5 &  0.50 & 1.8 & 1.2 &  0.40 & 2.5 & 0.32 &  0.11 \\
h016a & 54.00 & 2.6 & 3.5 &  0.57 & 5.0 & 1.3 &  0.21 & 2.3 & 4.30 &  0.65 \\ 
h016c & 51.00 & 0.081 & 12.0 &  0.99 & 0.27 & 11.0 &  0.98 & 0.027 & 12.00 &  1.00 \\ 
h019a & 57.00 & 0.45 & 4.3 &  0.91 & 0.45 & 5.0 &  0.92 & 0.45 & 9.0 &  0.67 \\ 
h019b & 56.00 & 0.050 & 2.6 &  0.98 & 0.32 & 2.7 &  0.90 & 0.0 & 1.70 &  1.00 \\ 
h087a & 112.00 & 0.0 & 86 &  1.00 & 2.8 & 0.94 &  0.25 & 0.072 & 59.00 &  1.00 \\
h087c & 119.00 & 0.16 & 5.8 &  0.97 & 0.32 & 6.6 &  0.95 & 2.2 & 0.019 &  0.01 \\ 
h088a & 80.00 & 14.0 & 9.9 &  0.42 & 14.0 & 11.0 &  0.45 & 7.8 & 18.00 &  0.70 \\ 
h088b & 80.00 & 12.0 & 18.0 &  0.61 & 14.0 & 14.0 &  0.50 & 3.9 & 24.00 &  0.86 \\ 
h088c & 80.00 & 1.0 & 5.4 &  0.84 & 1.1 & 5.2 &  0.82 & 0.069 & 5.30 &  0.99 \\ 
h088d & 80.00 & 1.2 & 4.2 &  0.78 & 1.2 & 4.2 &  0.78 & 1.7 & 0.09 &  0.05 \\
h089a & 117.00 & 2.2 & 15.0 &  0.87 & 2.5 & 15.0 &  0.86 & 0.31 & 18.00 &  0.98 \\ 
h089b & 120.00 & 0.84 & 2.6 &  0.76 & 1.6 & 2.0 &  0.56 & 1.3 & 2.6 &  0.67 \\ 
h089c & 117.00 & 1.1 & 9.1 &  0.90 & 1.4 & 11.0 &  0.88 & 0.012 & 15.00 &  1.00 \\ 
h091a & 92.50 & 1.1 & 6.3 &  0.85 & 2.2 & 5.8 &  0.72 & 0.11 & 3.0 &  0.96 \\ 
h091c & 97.60 & 2.8 & 0.43 &  0.13 & 3.1 & 0.084 &  0.03 & 3.2 & 0.002 &  0.00 \\ 
h096c & 118.00 & 0.45 & 0.40 &  0.47 & 0.45 & 0.54 &  0.55 & 0.31 & 0.52 &  0.63 \\ 

\enddata

\end{deluxetable}

\clearpage


\clearpage
\begin{figure*}

\includegraphics[scale=.8]{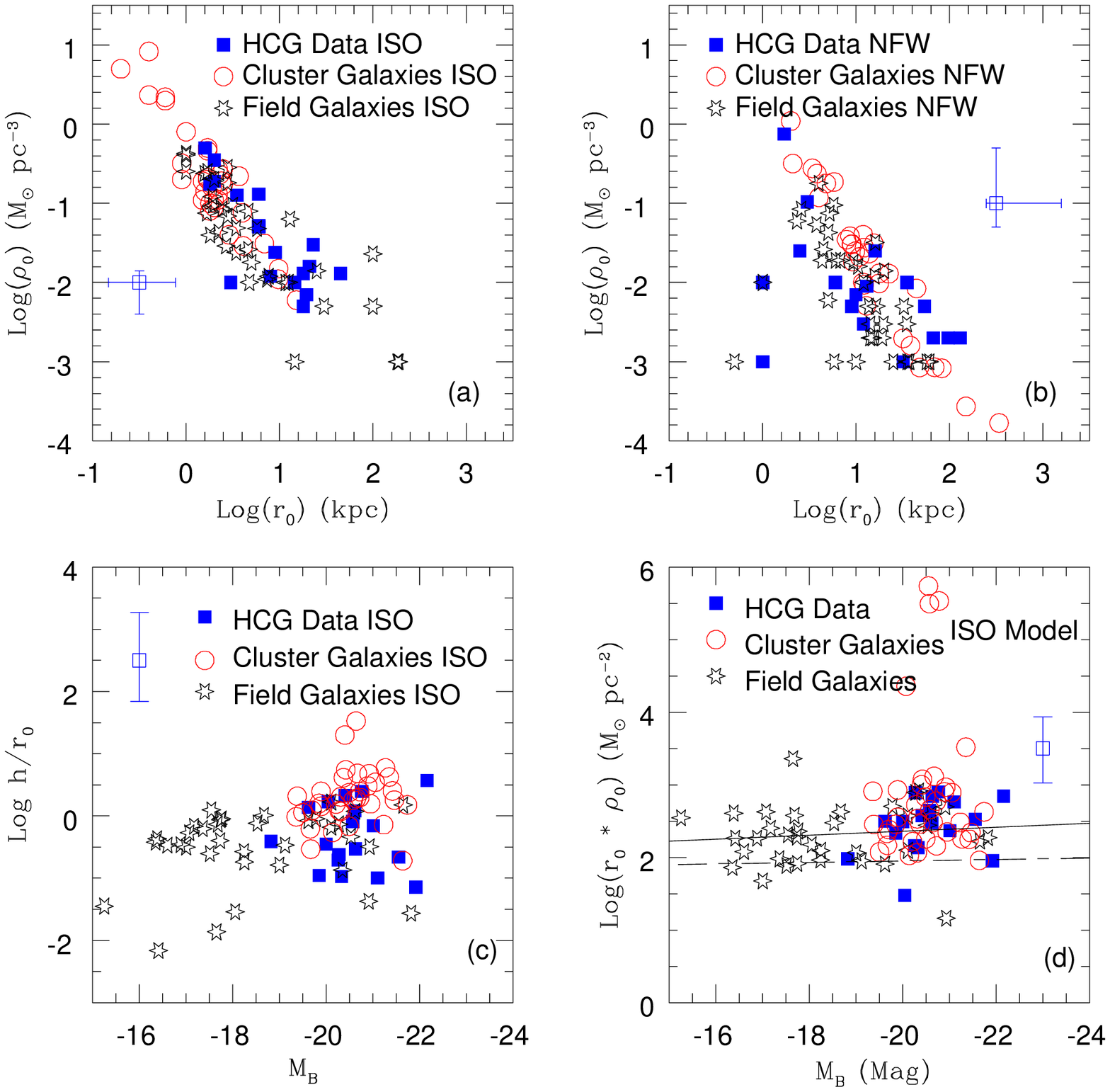}
\caption{Fig. 1a,b: Correlation between the central halo density and the halo core radius using the ISO model (a) and the NFW (b). Fig. 1c: Relation between the disk scale length normalised to the core radius vs the B absolute magnitude. Fig. 1d represents the halo central surface density vs the B absolute magnitude. The line represents the linear regression of the field galaxies and the dashed line represents the linear regression shown by \citep{Kor04}.  In each plot, error bars represent the maximum error found by the fit for each parameter. \label{Figure01}}
\end{figure*}

\clearpage


\begin{figure*}

\includegraphics[scale=0.95]{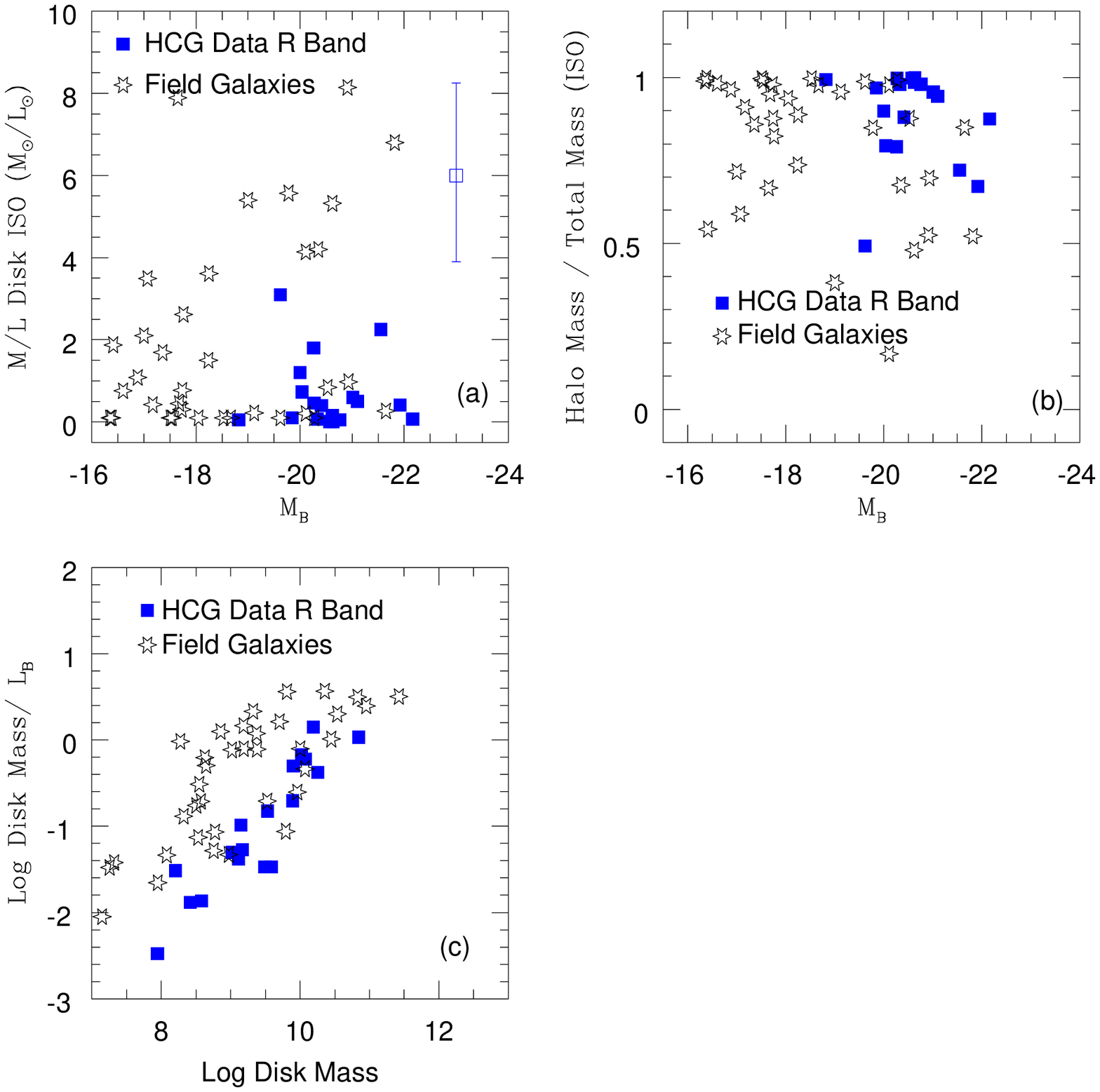}
\caption{Fig 2a represents the disk M/L using the ISO model vs the absolute B magnitude for both field and HCG galaxies. In each plot, error bars represent the maximum error found by the fit for each parameter.  Fig. 2b represents the halo mass fraction (normalised to the total mass) using ISO vs NFW for both field and HCG galaxies. Fig. 2c shows the disk mass divided by the total B luminosity vs the disk mass. \label{Figure02}}

\end{figure*}
\clearpage


\begin{figure*}
\vspace{-5.8cm}
\hspace{-2.0cm}
\includegraphics[scale=0.95]{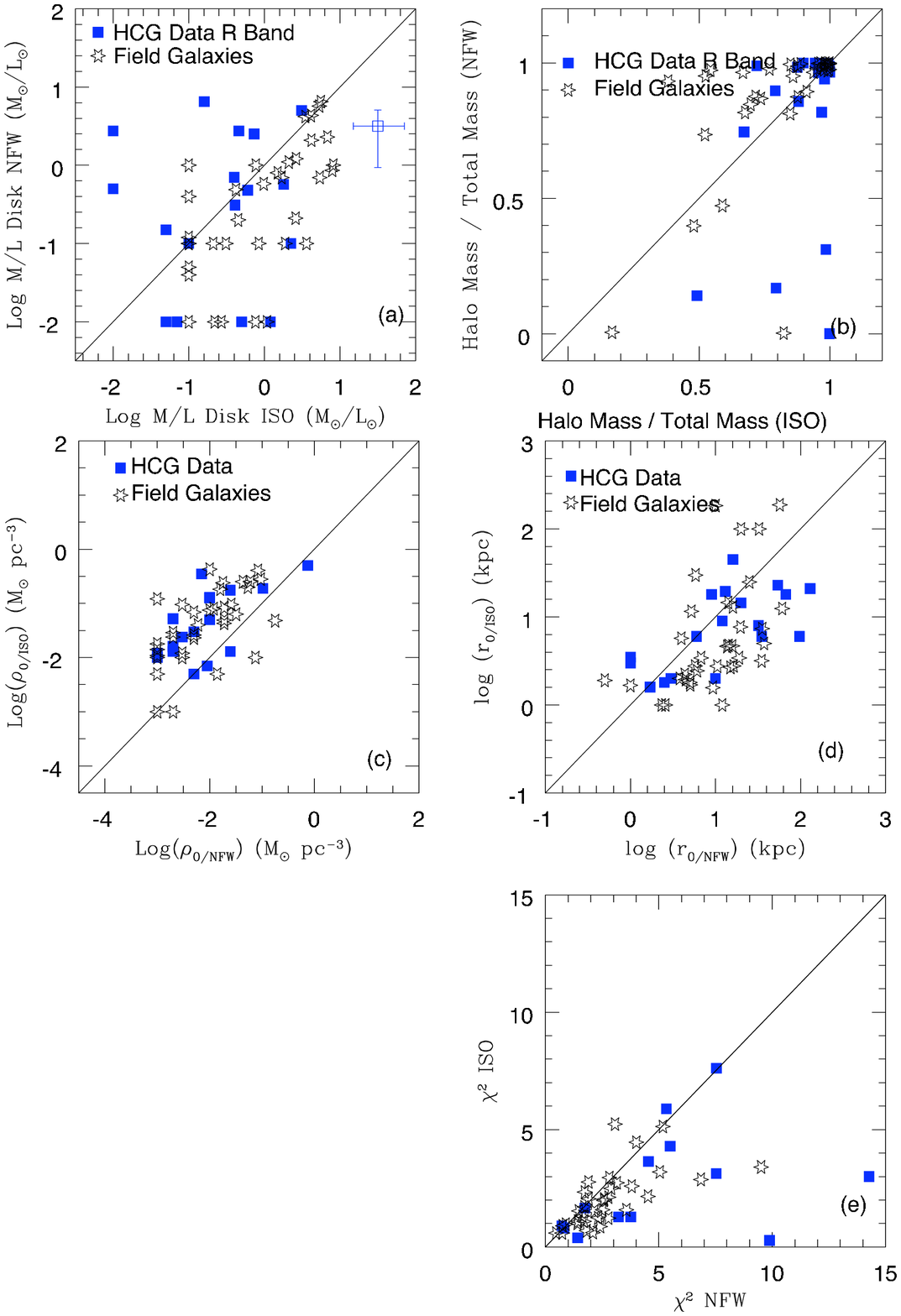}
\caption{Fig. 3a: Disk M/L using ISO vs NFW for both field and HCG galaxies.  Fig. 3b represents the halo mass fraction (normalised to the total mass) using ISO vs NFW for both field and HCG galaxies. Fig. 3c, d represent the comparison of the central halo density and the radius using ISO and NFW halo models. Fig 3e compares the reduced $\chi ^2$ parameter with ISO and NFW. Only 12 galaxies out 18 are shown; the others have reduced $\chi^2$  larger than 15. \label{Figure03}}
\end{figure*}




\end{document}